\documentclass[aps,prb,twocolumn,amsmath,amssymb,groupedaddress,longbibliography
]{revtex4-2}

\usepackage{graphicx}
\usepackage{dcolumn}
\usepackage{bm}
\usepackage{multirow}
\usepackage{braket}
\usepackage{color}
\usepackage{ulem}
\usepackage{booktabs}

\allowdisplaybreaks[1]

\begin{document}

\title{Essential model parameters for nonreciprocal magnons in multisublattice systems}

\author{Satoru Hayami and Takuya Matsumoto}
\affiliation{
Department of Applied Physics, the University of Tokyo, Tokyo 113-8656, Japan 
 }

\begin{abstract}
We theoretically investigate the microscopic conditions for emergent nonreciprocal
magnons toward unified understanding on the basis of a microscopic model analysis. 
We show that the products of the Bogoliubov Hamiltonian obtained within the linear spin wave approximation is enough to obtain the momentum-space functional form and the key ingredients in the nonreciprocal magnon dispersions in an analytical way even without solving the eigenvalue problems. 
We find that the odd order of an effective antisymmetric Dzyaloshinskii-Moriya interaction and/or the even order of an effective symmetric anisotropic interaction in the spin rotated frame can be a source of the antisymmetric dispersions. 
We present possible kinetic paths of magnons contributing to the antisymmetric dispersions in the one- to four-sublattice systems with the general exchange interactions.
We also test the formula for both ferromagnetic and antiferromagnetic orderings in the absence of spatial inversion symmetry. 
\end{abstract}

\maketitle

\section{Introduction}

Conductive phenomena in solids have long been studied in various fields of condensed matter physics, such as the giant magnetoresistance~\cite{baibich1988giant,ramirez1997colossal,tokura1999colossal,Tokura200005,dagotto2001colossal} and the anomalous Hall effect~\cite{Ye_PhysRevLett.83.3737,Ohgushi_PhysRevB.62.R6065,tatara2002chirality,Nagaosa_RevModPhys.82.1539,nagaosa2013topological,Lee_PhysRevLett.102.186601,nakatsuji2015large,Naka_PhysRevB.102.075112,Hayami_PhysRevB.103.L180407}. 
For these physical phenomena, the electronic band structures play an important role. 
The flat band structures give rise to magnetism, superconductivity, and the fractional quantum Hall effect~\cite{Tsui_PhysRevLett.48.1559,Lieb_PhysRevLett.62.1201,Kopnin_PhysRevB.83.220503,leykam2018artificial,Kuno_PhysRevB.102.241115,balents2020superconductivity}, while the linear band dispersions around the Dirac/Weyl points lead to unconventional topological properties~\cite{murakami2007phase,Wan_PhysRevB.83.205101,Burkov_PhysRevLett.107.127205,Young_PhysRevLett.108.140405,Armitage_RevModPhys.90.015001}. 
Besides, the spin splittings in the band structure bring about fascinating physical phenomena, such as the Edelstein effect in noncentrosymmetric systems~\cite{ivchenko1978new,edelstein1990spin,furukawa2017observation,yoda2018orbital}, spin current generation in antiferromagnetic systems without the relativistic spin-orbit coupling~\cite{Ahn_PhysRevB.99.184432,naka2019spin,hayami2019momentum,Yuan_PhysRevMaterials.5.014409}, and the spin-orbit-momentum locking in magnetic quadrupole systems~\cite{Hayami_PhysRevB.104.045117}. 

Under space-time inversion symmetry, the electronic band structures are categorized into four groups: the $\bm{k}$-symmetric band dispersion with the spin degeneracy in the presence of both spatial inversion ($\mathcal{P}$) and time-reversal ($\mathcal{T}$) symmetries, the $\bm{k}$-(anti)symmetric spin-split band dispersion without $\mathcal{T}$ ($\mathcal{P}$) while keeping $\mathcal{P}$ ($\mathcal{T}$), and the $\bm{k}$-antisymmetric band dispersion without both $\mathcal{P}$ and $\mathcal{T}$, where $\bm{k}$ is the wave vector of electrons. 
In particular, the $\bm{k}$-antisymmetric band dispersion has been extensively studied in recent years, since it becomes a source of nonreciprocal conductive phenomena owing to the inequivalence between $\bm{k}$ and $-\bm{k}$~\cite{tokura2018nonreciprocal}. 
The nonreciprocal nonlinear optical effect is a typical example~\cite{Sawada_PhysRevLett.95.237402,Toyoda_PhysRevB.93.201109,cheong2018broken,Foggetti_PhysRevB.100.180408}. 
The microscopic origin of the $\bm{k}$-antisymmetric band dispersion is accounted for by the active magnetic toroidal moment, which corresponds to a polar tensor with time-reversal odd~\cite{volkov1981macroscopic,kopaev2009toroidal,Spaldin_0953-8984-20-43-434203,Yanase_JPSJ.83.014703,Hayami_PhysRevB.90.024432,Hayami_doi:10.7566/JPSJ.84.064717,Watanabe_PhysRevB.98.220412,Hayami_PhysRevB.98.165110}. 

The nonreciprocal phenomena have also been discussed in magnetic insulators~\cite{damon1960magnetostatic,damon1961magnetostatic,Melcher_PhysRevLett.30.125,kataoka1987spin,PhysRevLett.57.2442,PhysRevB.35.5219,di2015enhancement,nembach2015linear,PhysRevB.93.235131,Otalora_PhysRevB.95.184415,Sasaki_PhysRevB.95.020407,garst2017collective,Weber_PhysRevB.97.224403,Cheon_PhysRevB.98.184405,tokura2018nonreciprocal,sato2019nonreciprocal,Nomura_PhysRevLett.122.145901,Lucassen_PhysRevB.101.064432,Santos_PhysRevB.102.104401, gallardo2019reconfigurable, PhysRevB.104.014402, zhang2015plane, mruczkiewicz2017spin}. 
In spite of the absence of carriers, the collective excitaions of magnons lead to directional-dependent dynamical properties, where we refer it to the nonreciprocal (asymmetric) magnons~\cite{tokura2018nonreciprocal,sato2019nonreciprocal}. 
Similar to the electron band dispersion, an appearance of nonreciprocal magnons is attributed to the active magnetic toroidal moment~\cite{Hayami_doi:10.7566/JPSJ.85.053705}. 
Although they were mainly studied for ferromagnetic slabs~\cite{damon1960magnetostatic,damon1961magnetostatic} and for magnetic orderings in the noncentrosymmetric crystals~\cite{Melcher_PhysRevLett.30.125,kataoka1987spin,garst2017collective,kawano2019designing,matsumoto2021nonreciprocal}, where the magnetic dipolar interaction and/or the Dzyaloshinskii-Moriya (DM) interaction are important~\cite{dzyaloshinsky1958thermodynamic,moriya1960anisotropic}, it was shown that they occur even via other mechanisms, such as frustrated exchange interactions~\cite{Miyahara_JPSJ.81.023712,Miyahara_PhysRevB.89.195145} and bond-dependent symmetric exchange interactions~\cite{Maksimov_PhysRevX.9.021017,Matsumoto_PhysRevB.101.224419}. 
The nonreciprocal magnons have a potential to exhibit further intriguing nonreciprocal phenomena, such as the magneto-optical effect~\cite{Mochizuki_PhysRevLett.114.197203,Proskurin_PhysRevB.98.134422,Okuma_PhysRevB.99.094401} and spin Seebeck effect~\cite{Takashima_PhysRevB.98.020401}, which avoid Joule heating.

Engineering asymmetric band deformations in the systems without $\mathcal{P}$ and $\mathcal{T}$ symmetries is important for nonreciprocal conductive phenomena irrespective of electrons and magnons. 
Meanwhile, the microscopic conditions have not been fully clarified yet, although active magnetic toroidal multipoles are necessary from the symmetry aspect~\cite{hayami2018microscopic,Hayami_PhysRevB.98.165110,kusunose2020complete,Yatsushiro_PhysRevB.104.054412}. 
Recently, a useful framework to extract essential model parameters for the asymmetric band structure in the electron systems has been proposed on the basis of augmented multipoles~\cite{Hayami_PhysRevB.102.144441}. 
Similar approach has also been performed in the magnon systems by introducing the bond-type magnetic toroidal dipole degree of freedom, which is only applied to the mechanism induced by the DM interaction~\cite{matsumoto2021nonreciprocal}. 
It is desired to have a simple formula to investigate which model parameters contribute to the asymmetric band deformations in magnon systems with arbitrary spin interactions. 

In the present study, we investigate the microscopic conditions for emergent nonreciprocal
magnons in multisublattice systems in an analytical way. 
We show that the product of the Bogoliubov Hamiltonian after the linear spin wave approximation provides two important information for nonreciprocal magnons without the cumbersome Bogoliubov transformation. 
One is the momentum-space functional form and the other is the essential model parameters to cause the antisymmetric band deformations. 
We demonstrate that our scheme ubiquitously accounts for the microscopic key ingredients irrespective of the mechanisms by analyzing a spin Hamiltonian with the general exchange interactions in the one- to four-sublattice systems. 
We discuss the important magnon-hopping processes that arise from the exchange interactions in real space. 
We also test our scheme for both ferromagnetic and antiferromagnetic orderings with the DM interaction and the symmetric anisotropic interaction. 
Our results will be useful to extract the significant model parameters in inducing the nonreciprocal magnons under complicated noncollinear magnetic orderings.

The remaining of the paper is organized as follows. 
In Sec.~\ref{sec:Approach}, we present a general method of extracting the essential model parameters from the Bogoliubov Hamiltonian. 
We present a general expression contributing to nonreciprocal magnons on the basis of the spin Hamiltonian with both symmetric and antisymmetric exchange interactions in the one- to four-sublattice systems in Sec.~\ref{sec:General feature of F}.
We apply the method for the ferromagnetic ordering in the breathing kagome lattice structure and the collinear/noncollinear antiferromagnetic orderings in the honeycomb and breathing kagome lattice structures in Sec.~\ref{sec:Application to noncentrosymmetric magnets}. 
Section~\ref{sec:Summary} is devoted to a summary of the present paper. 
Appendix~\ref{sec:appendix} provides lengthy expressions in terms of momentum-space functions in the three- and four-sublattice cases.

\section{Approach}
\label{sec:Approach}
Let us start a general spin Hamiltonian, which is given by 
\begin{align}
\label{eq:Ham_init}
H=\sum_{ll'} \sum_{\alpha \beta} S^{\alpha}_l \mathcal{J}^{\alpha \beta}_{ll'}
S^{\beta}_{l'}, 
\end{align}
with 
\begin{align}
\label{eq:Jmat}
\mathcal{J}_{ll'}=\left(
\begin{array}{ccc}
J_{ll'}^{\perp}+J_{ll'}^v & J_{ll'}^{xy}+D_{ll'}^{z} & J_{ll'}^{zx}-D_{ll'}^{y} \\
J_{ll'}^{xy}-D_{ll'}^{z}& J_{ll'}^{\perp}-J_{ll'}^v & J_{ll'}^{yz}+D_{ll'}^{x} \\ 
J_{ll'}^{zx}+D_{ll'}^{y}& J_{ll'}^{yz}-D_{ll'}^{x} & J_{ll'}^{z} \\ 
\end{array}
\right),  
\end{align}
where $S^{\alpha}_{l}$ is an $\alpha$ ($=x$, $y$, and $z$) component of classical spin at site $l$.
 $J^{\perp}_{ll'}$, $J^{z}_{ll'}$, $J^{v}_{ll'}$, $J^{xy}_{ll'}$, $J^{yz}_{ll'}$, and $J^{zx}_{ll'}$ are the symmetric exchange interactions, while $D^{x }_{ll'}$, $D^{y }_{ll'}$, and $D^{z}_{ll'}$ are the antisymmetric exchange interactions. The latter corresponds to the DM interaction. 
The nonzero components of $\mathcal{J}_{ll'}$ are determined by point group symmetry of the bond. 
For later convenience, the spin is rotated so as to align the local axis along the $z$ direction: 
\begin{align}
\label{eq:spinrotate}
(S_l^x, S_l^y, S_l^z )^{\rm T} = R_z (\phi_l)R_y(\theta_l) (\tilde{S}_l^x,\tilde{S}_l^y,\tilde{S}_l^z )^{\rm T}, 
\end{align}
where $R_z (\phi_l)$ and $R_{y}(\theta_l)$ are the rotation matrices around the $z$ and $y$ axes, respectively, and T is the transpose of the vector. 
Then, the Hamiltonian in Eq.~(\ref{eq:Ham_init}) is rewritten as 
\begin{align}
\label{eq:Ham_init_trans}
H&=\sum_{ll'} \sum_{\alpha \beta} \tilde{S}^{\alpha}_l \mathcal{\tilde{J}}_{ll'} ^{\alpha \beta}\tilde{S}^{\beta}_{l'} \nonumber \\
&=\sum_{ll'} \left(H^{\perp}_{ll'}+H^{\rm {DM}}_{ll'}+H^{v}_{ll'}+H^{xy}_{ll'}+H^{z}_{ll'}+H^{yz/zx}_{ll'}\right), 
\end{align}
where
\begin{align}
\label{eq:rot_J}
H^{\perp}_{ll'}&= \frac{\tilde{J}_{ll'}^{\perp} }{2}(\tilde{S}^+_{l} \tilde{S}^-_{l'}+ \tilde{S}^-_{l} \tilde{S}^+_{l'}), \\
\label{eq:rot_D}
H^{\rm {DM}}_{ll'}&=\frac{ i \tilde{D}_{ll'} }{2}(\tilde{S}^+_{l} \tilde{S}^-_{l'}- \tilde{S}^-_{l} \tilde{S}^+_{l'}), \\
\label{eq:rot_v}
H^{v}_{ll'}&= \frac{\tilde{J}_{ll'}^{v} }{2}(\tilde{S}^+_{l}\tilde{S}^+_{l'}+\tilde{S}^-_{l} \tilde{S}^-_{l'}), \\
\label{eq:rot_xy}
H^{xy}_{ll'}&=-\frac{ i \tilde{J}_{ll'}^{xy} }{2}(\tilde{S}^+_{l} \tilde{S}^+_{l'}-\tilde{S}^-_{l} \tilde{S}^-_{l'}), \\
\label{eq:rot_z}
H^{z}_{ll'}&= \tilde{J}^{z}_{ll'} \tilde{S}^z_{l}\tilde{S}^z_{l'}.  
\end{align}
$H^{zx}_{ll'}$ and $H^{yz}_{ll'}$ consist of the product of $\tilde{S}^x \tilde{S}^z$ and $\tilde{S}^y \tilde{S}^z$,
 respectively. 
The interaction tensor $\mathcal{\tilde{J}}_{ll'}$ is represented by rotating $\mathcal{J}_{ll'}$. 

We investigate magnon spectra within a linear spin wave approximation.  
By applying the Holstein-Primakov transformation, which is given by $\tilde{S}^+_{i\eta}=\sqrt{2S}a_{i\eta}$, $\tilde{S}^-_{i\eta}=\sqrt{2S}a_{i\eta}^{\dagger}$, and $\tilde{S}^z_{i\eta}=S-a_{i\eta}^{\dagger} a_{i\eta}$ (the subscripts $i$ and $\eta$ denote the indices for a unit cell and a sublattice, respectively, and $a_{i\eta}$ is the boson operator for sublattice $\eta$), to the spin Hamiltonian in Eq.~(\ref{eq:Ham_init_trans}), the Bogoliubov Hamiltonian is derived. 
By performing the Fourier transformation as $a_{i\eta} \to a_{\bm{q}\eta}$, the resultant Bogoliubov Hamiltonian in the $n$-sublattice system is given by 
\begin{align}
\label{eq:Ham_B}
H^{\rm B}&=\frac{S}{2}\sum_{\bm{q}} \Psi^{\dagger}_{\bm{q}} H^{\rm B}_{\bm{q}}
\Psi_{\bm{q}}, \\
\label{eq:Ham_B2}
H^{\rm B}_{\bm{q}}&=\left(
\begin{array}{cc}
\mathcal{X}_{\bm{q}} & \mathcal{Y}_{\bm{q}} \\
\mathcal{Y}^\dagger_{\bm{q}} & \mathcal{X}^*_{-\bm{q}}
\end{array}
\right), 
\end{align}
where $\Psi^{\dagger}_{\bm{q}}=(a^{\dagger}_{\bm{q}1}, a^{\dagger}_{\bm{q}2}, \cdots, a^{\dagger}_{\bm{q}n},a_{-\bm{q}1}, a_{-\bm{q}2}, \cdots, a_{-\bm{q}n})$ and $\mathcal{X}_{\bm{q}}$ and $\mathcal{Y}_{\bm{q}}$ are the $n\times n$ matrices. 

In Eq.~(\ref{eq:Ham_init_trans}), $H^{z}_{ll'}$ corresponds to the diagonal elements of $\mathcal{X}_{\bm{q}}$, while $H^{\perp}_{ll'}$, $H^{\rm {DM}}_{ll'}$, $H^{v}_{ll'}$, and $H^{xy}_{ll'}$ correspond to the off-diagonal elements $\mathcal{X}_{\bm{q}}$ and $\mathcal{Y}_{\bm{q}}$. 
In other words, only the spin components perpendicular to $\tilde{S}^z_{l}$ contribute to a magnon hopping process. 
Meanwhile, $H^{yz/zx}_{ll'}$ does not appear in Eq.~(\ref{eq:Ham_B}), since it consists of the odd number of boson operators.

When $H^{\rm B}_{\bm{q}}$ is a positive-definite matrix, the Cholesky decomposition is possible as $H^{\rm B}_{\bm{q}}=K^\dagger_{\bm{q}}K_{\bm{q}}$, where $K_{\bm{q}}$ is  the upper triangular matrix. 
Then, $H^{\rm B}_{\bm{q}}$ is transformed into the Hermitian matrix $H_{\bm{q}}$ as 
\begin{align}
H_{\bm{q}}=K_{\bm{q}}g K^\dagger_{\bm{q}}, 
\end{align}
where the $2n \times 2n$ matrix $g$ satisfies $(g)_{\eta\eta'}=[\Psi_{\bm{q}\eta} ,\Psi^{\dagger}_{\bm{q}\eta'} ]$. 
The eigenvalues $\omega_{\bm{q}m}$ ($m$ is the band index) in Eq.~(\ref{eq:Ham_B2}) are obtained by diagonalizing $H_{\bm{q}}$.

Nonreciprocal magnon excitations mean that the eigenvalues have an antisymmetric component with respect to $\bm{q}$, i.e., $\omega_{\bm{q}m} \neq \omega_{-\bm{q}m}$. 
To investigate important model parameters for the nonreciprocal magnons in a systematic way, we introduce a following quantity as 
\begin{align}
E^{(s)}_{\bm{q}}&={\rm Tr}[\underbrace{H_{\bm{q}}H_{\bm{q}} \cdots H_{\bm{q}}}_{s} ],  \\
\label{eq:Es}
&={\rm Tr}[\underbrace{(H^{\rm B}_{\bm{q}}g)( H^{\rm B}_{\bm{q}}g) \cdots (H^{\rm B}_{\bm{q}}g}_s)],
\end{align}
which is related to the eigenenergy. 
A similar quantity has been discussed in the antisymmetric band modulation and spin splittings in the electron system~\cite{Hayami_PhysRevB.101.220403,Hayami_PhysRevB.102.144441,oiwa2021systematic}. 
The antisymmetric component is extracted by 
\begin{align}
\label{eq:Fs}
F^{(s)}_{\bm{q}}&=\frac{1}{2}(E^{(s)}_{\bm{q}}-E^{(s)}_{-\bm{q}}). 
\end{align}
Thus nonzero $F^{(s)}_{\bm{q}}$ signals the appearance of nonreciprocal magnons.

From the expression of Eq.~(\ref{eq:Es}), one can deduce the essential model parameters inducing nonreciprocal magnons, as detailed in Sec.~\ref{sec:General feature of F}. 
In Eqs.~(\ref{eq:rot_J})-(\ref{eq:rot_z}), there are four types of magnon hoppings and one onsite potential in the real space Bogoliubov Hamiltonian, which are expressed as
\begin{align}
\label{eq:rot_J_magnon}
H^{\perp}_{ll'}&=
 S \tilde{J}_{ll'}^{\perp}
(a_{l} a^\dagger_{l'} + a_l^\dagger a_{l'})
, \\
\label{eq:rot_D_magnon}
H^{\rm {DM}}_{ll'}&=
  i S\tilde{D}_{ll'} (a_l a^\dagger_{l'}- a_l^\dagger a_{l'}), \\
\label{eq:rot_v_magnon}
H^{v}_{ll'}&= 
S \tilde{J}_{ll'}^{v} (a_l a_{l'} + a_{l}^\dagger a_{l'}^\dagger), \\
\label{eq:rot_xy_magnon}
H^{xy}_{ll'}&= - iS \tilde{J}_{ll'}^{xy} (a_{l} a_{l'} -a^\dagger_{l} a^\dagger_{l'}), \\
\label{eq:rot_z_magnon}
H^{z}_{ll'}&= S \tilde{J}^{z}_{ll'} (S - a_l^\dagger a_{l}-a_{l'}^\dagger a_{l'}). 
\end{align}
From theses expressions, one finds that the real (imaginary) part of the standard hopping $a^\dagger_{i\eta} a_{j\eta'}$ is related to $H^{\perp}_{ll'}$ ($H^{\rm {DM}}_{ll'}$), which corresponds to the off-diagonal part of $\mathcal{X}_{\bm{q}}$, while the real (imaginary) part of the anomalous hopping $a_{i\eta}^\dagger a^\dagger_{j\eta'}$ is related to $H^{v}_{ll'}$ ($H^{xy}_{ll'}$), which corresponds to the off-diagonal part of $\mathcal{Y}_{\bm{q}}$. 
As only the hopping processes to satisfy the magnon-number conservation are important, one can find that an even order of $\tilde{J}^v_{ll'}$ and $\tilde{J}^{xy}_{ll'}$ can contribute to nonreciprocal magnon excitations. 
In addition, when taking into account the fact that an odd order of imaginary hopping can also contribute to nonreciprocal magnon excitations, we expect that the antisymmetric magnon band structure is related to the odd order of an effective antisymmetric DM interaction or the even order of an effective symmetric anisotropic interaction.
This indicates that the antisymmetric magnon band structure can be reversed regarding $\bm{q}$ by the sign of $\tilde{D}_{ll'}$, while that is not by the sign of $\tilde{J}^v_{ll'}$ and $\tilde{J}^{xy}_{ll'}$. 
As we will show the general feature of $F^{(s)}_{\bm{q}}$ in Sec.~\ref{sec:General feature of F} and the specific examples in Sec.~\ref{sec:Application to noncentrosymmetric magnets}, the quantity $F^{(s)}_{\bm{q}}$ gives a microscopic condition of nonreciprocal magnons irrespective of ferromagnets and antiferromagnets. 

\section{General feature of $F^{(s)}_{\bm{q}}$}
\label{sec:General feature of F}

In this section, we discuss a general behavior of $F^{(s)}_{\bm{q}}$ independent of the lattice structures and the exchange interactions. 
We show the microscopic processes contributing to nonreciprocal magnons in the multisublattice systems with $n=1$-4: one-sublattice case in Sec.~\ref{sec:One-sublattice case}, two-sublattice case in Sec.~\ref{sec:Two-sublattice case}, three-sublattice case in Sec.~\ref{sec:Three-sublattice case}, and four-sublattice case in Sec.~\ref{sec:Four-sublattice case}. 
It is noted that the present scheme can be also applied to the systems with the sublattice $n>4$ in a straightforward way.

\subsection{One-sublattice case}
\label{sec:One-sublattice case}

\begin{figure}[t!]
\begin{center}
\includegraphics[width=0.3 \hsize]{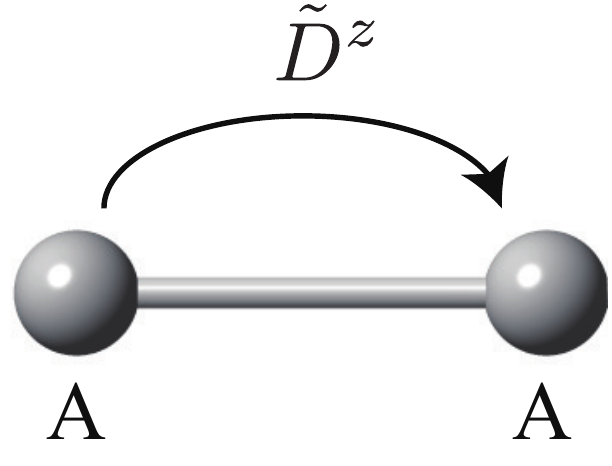} 
\caption{
\label{Fig:1sub}
Schematic picture of the magnon-hopping process contributing to nonreciprocal magnons ($F^{(1)}_{\bm{q}} \neq 0$) in real space in the one-sublattice case. 
}
\end{center}
\end{figure}

We consider the one-sublattice system with $\eta=$ A, which describes only the ferromagnetic state without the sublattice degree of freedom. 
In the one-sublattice system, $\mathcal{X}_{\bm{q}}$ and $\mathcal{Y}_{\bm{q}}$ are the $1\times 1$ matrices. 
By using Eqs.~(\ref{eq:rot_J_magnon})-(\ref{eq:rot_z_magnon}), the expressions of $\mathcal{X}_{\bm{q}}$ and $\mathcal{Y}_{\bm{q}}$ are given by 
\begin{align}
\mathcal{X}_{\bm{q}}&=\tilde{J}^z h^{z{\rm (s)}}_{\bm{q}} + \tilde{J}^{\perp} h^{\perp{\rm (s)}}_{\bm{q}}-\tilde{D}^z h^{D{\rm (as)}}_{\bm{q}},\\
\mathcal{Y}_{\bm{q}}&= \tilde{J}^{v}
h^{v{\rm (s)}}_{\bm{q}}
+ i\tilde{J}^{xy} 
h^{xy{\rm (s)}}_{\bm{q}}, 
\end{align}
where $h^{\zeta{\rm (s)}}_{\bm{q}}$ and $h^{\zeta{\rm (as)}}_{\bm{q}}$ for $\zeta=z, \perp, D, v, xy$ are arbitrary symmetric and antisymmetric functions with respect to $\bm{q}$: $h^{\zeta{\rm (s)}}_{\bm{q}}=h^{\zeta{\rm (s)}}_{-\bm{q}}$ and $h^{\zeta{\rm (as)}}_{\bm{q}}=-h^{\zeta{\rm (as)}}_{-\bm{q}}$.
Owing to the one-sublattice degree of freedom, $h^{\perp{\rm (as)}}_{\bm{q}}=h^{D{\rm (s)}}_{\bm{q}}=h^{v{\rm (as)}}_{\bm{q}}=h^{xy{\rm (as)}}_{\bm{q}}=0$ and $h^{z{\rm (s)}}_{\bm{q}}$ has a $\bm{q}$ dependence, which are different from the multisublattice cases, as will be discussed in Secs.~\ref{sec:Two-sublattice case}-\ref{sec:Four-sublattice case}.

Although the magnon dispersions in the one-sublattice case with the $2\times 2$ matrix $H^{\rm B}_{\bm{q}}$ are analytically obtained by performing the Bogoliubov transformation, we test the expressions in Eqs.~(\ref{eq:Es}) and (\ref{eq:Fs}) for later complicated multisublattice systems.
The lowest contribution of $F^{(s)}_{\bm{q}}$ is given by 
\begin{align}
\label{eq:Fs_1sub_D}
F^{(1)}_{\bm{q}}&= -2 \tilde{D}^z h^{D{\rm (as)}}_{\bm{q}}. 
\end{align}
The expression in Eq.~(\ref{eq:Fs_1sub_D}) indicates that only the effective DM interaction $\tilde{D}^z$ contributes to nonreciprocal magnon dispersions. 
When calculating the higher order of $F^{(s)}_{\bm{q}}$, one finds that the $(2m+1)$th-order terms of $F^{(s)}_{\bm{q}}$ are proportional to $\tilde{D}^z h^{D{\rm (as)}}_{\bm{q}}$, while the $2m$th-order ones 
vanish for an integer $m$. 
This means that the nonreciprocal magnon in the one-sublattice system is induced when $\tilde{D}^z\neq 0$ irrespective of other interactions.
This result is consistent with that obtained by the direct diagonalization.

The above result is intuitively understood from the magnon-hopping process in the real-space picture, as shown in the case of $F^{(1)}_{\bm{q}}$ in Fig.~\ref{Fig:1sub}. 
The process in Fig.~\ref{Fig:1sub} gives rise to effective imaginary magnon hopping that is a source of nonreciprocal magnons along the hopping direction. 
Furthermore, the functional form of nonreciprocal magnons are obtained in an analytic form from Eq.~(\ref{eq:Fs_1sub_D}).  
In the crystal system, the $\bm{q}$ dependence of $F^{(s)}_{\bm{q}}$ is derived to satisfy the magnetic point group symmetry in the system, as shown in Sec.~\ref{sec:Application to noncentrosymmetric magnets}.

\subsection{Two-sublattice case}
\label{sec:Two-sublattice case}

\begin{figure}[t!]
\begin{center}
\includegraphics[width=1.0 \hsize]{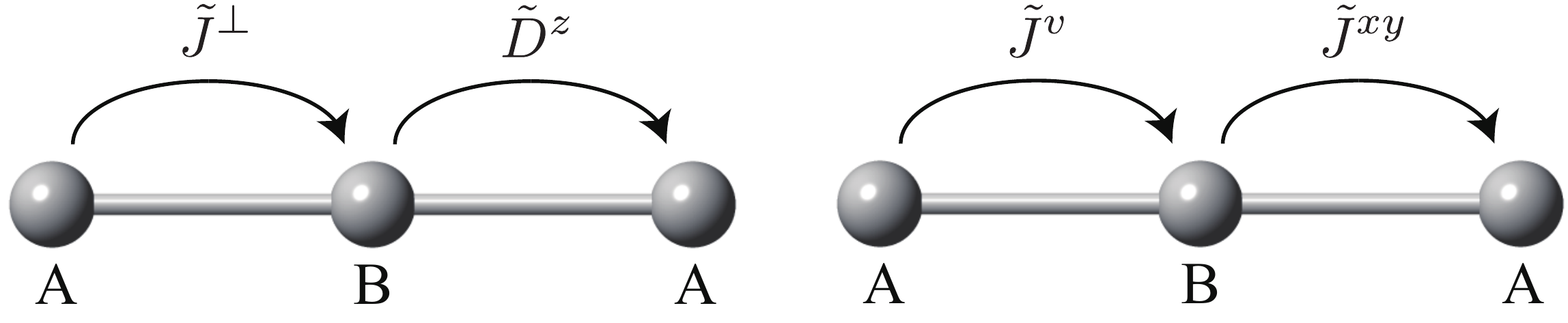} 
\caption{
\label{Fig:2sub}
Schematics of two magnon-hopping processes giving $F^{(3)}_{\bm{q}} \neq 0$ in real space in the two-sublattice case. 
The left panel corresponds to the first term in Eq.~(\ref{eq:Fs_2sub}) and the right panel corresponds to the second term in Eq.~(\ref{eq:Fs_2sub}). 
}
\end{center}
\end{figure}

Hereafter, we examine $F^{(s)}_{\bm{q}}$ in the multisublattice case. 
In this section, we show $F^{(s)}_{\bm{q}}$ in the two-sublattice case with $\eta=$ A and B, where $\mathcal{X}_{\bm{q}}$ and $\mathcal{Y}_{\bm{q}}$ are the $2\times 2$ matrices. 
By considering the general exchange interactions between A and B sublattices, $\mathcal{X}_{\bm{q}}$ and $\mathcal{Y}_{\bm{q}}$ are represented by 
\begin{align}
\label{eq:HL_X_2sub}
\mathcal{X}_{\bm{q}}=&
\begin{pmatrix}
Z_{{\rm A}}&F_{{\rm AB}\bm{q}}\\
F^*_{{\rm AB}\bm{q}}&Z_{{\rm B}}
\end{pmatrix},\\
\label{eq:HL_Y_2sub}
\mathcal{Y}_{\bm{q}}=&
\begin{pmatrix}
 0&G_{{\rm AB}\bm{q}}\\
G_{{\rm AB}-\bm{q}}&0
\end{pmatrix},
\end{align}
where 
\begin{align}
F_{{\rm AB}\bm{q}}&=\tilde{J}^{\perp} (h^{\perp{\rm (s)}}_{{\rm AB}\bm{q}}+i h^{\perp{\rm (as)}}_{{\rm AB}\bm{q}})
+i \tilde{D}^z (h^{D{\rm (s)}}_{{\rm AB}\bm{q}}+i h^{D{\rm (as)}}_{{\rm AB}\bm{q}}),\\
G_{{\rm AB}\bm{q}}&=\tilde{J}^{v} (h^{v{\rm (s)}}_{{\rm AB}\bm{q}}+i h^{v{\rm (as)}}_{{\rm AB}\bm{q}})
+ i\tilde{J}^{xy} (h^{xy{\rm (s)}}_{{\rm AB}\bm{q}}+i h^{xy{\rm (as)}}_{{\rm AB}\bm{q}}), \\
Z_{{\eta}}&=J^z z_{\eta}, 
\end{align}
and $\eta=$ A and B. 
In contrast to the one-sublattice case, $h^{\perp{\rm (as)}}_{\bm{q}} \neq 0$, $h^{D{\rm (s)}}_{\bm{q}} \neq 0$, $h^{v{\rm (as)}}_{\bm{q}} \neq 0$, and $h^{xy{\rm (as)}}_{\bm{q}} \neq 0$ and there is no $\bm{q}$ dependence in $Z_{\eta}$; $h^{z{\rm (s)}}_{\bm{q}}$ corresponds to $z_{\eta}$ and $h^{z{\rm (as)}}_{\bm{q}}=0$.

The lowest contribution of $F^{(s)}_{\bm{q}}$ is given by $s=3$, whose expression is represented as 
\begin{align}
\label{eq:Fs_2sub}
F^{(3)}_{\bm{q}}&= 12 \tilde{J}^z \tilde{D}^{z} \tilde{J}^{\perp} (z_{\rm A}+z_{\rm B})
(h^{D{\rm (s)}}_{{\rm AB}\bm{q}} h^{\perp{\rm (as)}}_{{\rm AB}\bm{q}} - h^{\perp{\rm (s)}}_{{\rm AB}\bm{q}} h^{D{\rm (as)}}_{{\rm AB}\bm{q}} )
 \nonumber \\
&-12 \tilde{J}^z\tilde{J}^{v} \tilde{J}^{xy} (z_{\rm A}-z_{\rm B}) (h^{xy{\rm (s)}}_{{\rm AB}\bm{q}} h^{v{\rm (as)}}_{{\rm AB}\bm{q}}-h^{v{\rm (s)}}_{{\rm AB}\bm{q}} h^{xy{\rm (as)}}_{{\rm AB}\bm{q}}). 
\end{align}
The first term in Eq.~(\ref{eq:Fs_2sub}) represents the contribution from the effective DM interaction proportional to $\tilde{D}^{z}$, which is similar to the result in the one-sublattice case in Sec.~\ref{sec:One-sublattice case}.
Meanwhile, the second term in Eq.~(\ref{eq:Fs_2sub}) represents the contribution from the effective symmetric anisotropic exchange interaction including $\tilde{J}^{v}$ and $\tilde{J}^{xy}$, which does not appear in the one-sublattice case. 
In other words, the symmetric anisotropic exchange interaction can become a source of nonreciprocal magnons in the multisublattice system [see also the results in Eq.~(\ref{eq:Fs_3sub}) in the three-sublattice case (Sec.~\ref{sec:Three-sublattice case}) and in Eq.~(\ref{eq:Fs_4sub}) in the four-sublattice case (Sec.~\ref{sec:Four-sublattice case})]. 
The real-space pictures in terms of the magnon-hopping processes for each term are shown in Fig.~\ref{Fig:2sub}. 
It is noted that the effective symmetric anisotropic interaction contributes to the nonreciprocal magnons in the form of $\tilde{J}^v \tilde{J}^{xy}$ in order to satisfy the magnon-number conservation and the space-time inversion symmetry.
We also note that the $\bm{q}$ dependence of nonreciprocal magnons can be different for different mechanisms, as found in the first and second terms in Eq.~(\ref{eq:Fs_2sub}).

In addition, there are three differences from the one-sublattice case in Eq.~(\ref{eq:Fs_1sub_D}). 
The one is the appearance of $\tilde{J}^z$ in Eq.~(\ref{eq:Fs_2sub}), which means that $\tilde{J}^z$ is also important to induce the nonreciprocal magnons. 
The second is the sublattice-dependent factor $z_{\rm A}+z_{\rm B}$ and $z_{\rm A}-z_{\rm B}$; the nonreciprocal magnons by $\tilde{D}^{z}$ ($\tilde{J}^{v} \tilde{J}^{xy}$) vanish when $z_{\rm A}=-z_{\rm B}$ ($z_{\rm A}=z_{\rm B}$). 
The third is the $\bm{q}$ dependence in the first term in Eq.~(\ref{eq:Fs_2sub}) owing to nonzero $h^{\perp{\rm (as)}}_{\bm{q}}$ and $h^{D{\rm (s)}}_{\bm{q}}$.

We note that the expression in Eq.~(\ref{eq:Fs_2sub}) does not directly reduce to that in Eq.~(\ref{eq:Fs_1sub_D}) when regarding A and B sublattices as the same sublattice, i.e., $z_{\rm A}=z_{\rm B}$: The essential model parameter in Eq.~(\ref{eq:Fs_2sub}) is $\tilde{J}^z \tilde{D}^{z} \tilde{J}^{\perp}$, while that in Eq.~(\ref{eq:Fs_1sub_D}) is $\tilde{D}^{z}$. 
At first glance this result appears to contradict with each other, but it is due to the fact that 
the factor $\tilde{J}^z \tilde{J}^{\perp}$ is canceled out with the denominators when evaluating the energy spectrum~\cite{matsumoto2021nonreciprocal}. 
Hence, from the viewpoint of obtaining the essential model parameters, it is useful to calculate $F^{(s)}_{\bm{q}}$ in the minimal unit cell.

By using the expression in Eq.~(\ref{eq:Fs_2sub}), one obtains the essential model parameters for the emergence of nonreciprocal magnons in the two-sublattice antiferromagnetic orderings and the ferromagnetic ordering in the two-sublattice noncentrosymmetric structures.  
We show the example of the staggered antiferromagnetic ordering in the honeycomb lattice structure in Sec.~\ref{sec:Nonreciprocal magnon in honeycomb antiferromagnets}.

\subsection{Three-sublattice case}
\label{sec:Three-sublattice case}

\begin{figure}[t!]
\begin{center}
\includegraphics[width=0.8 \hsize]{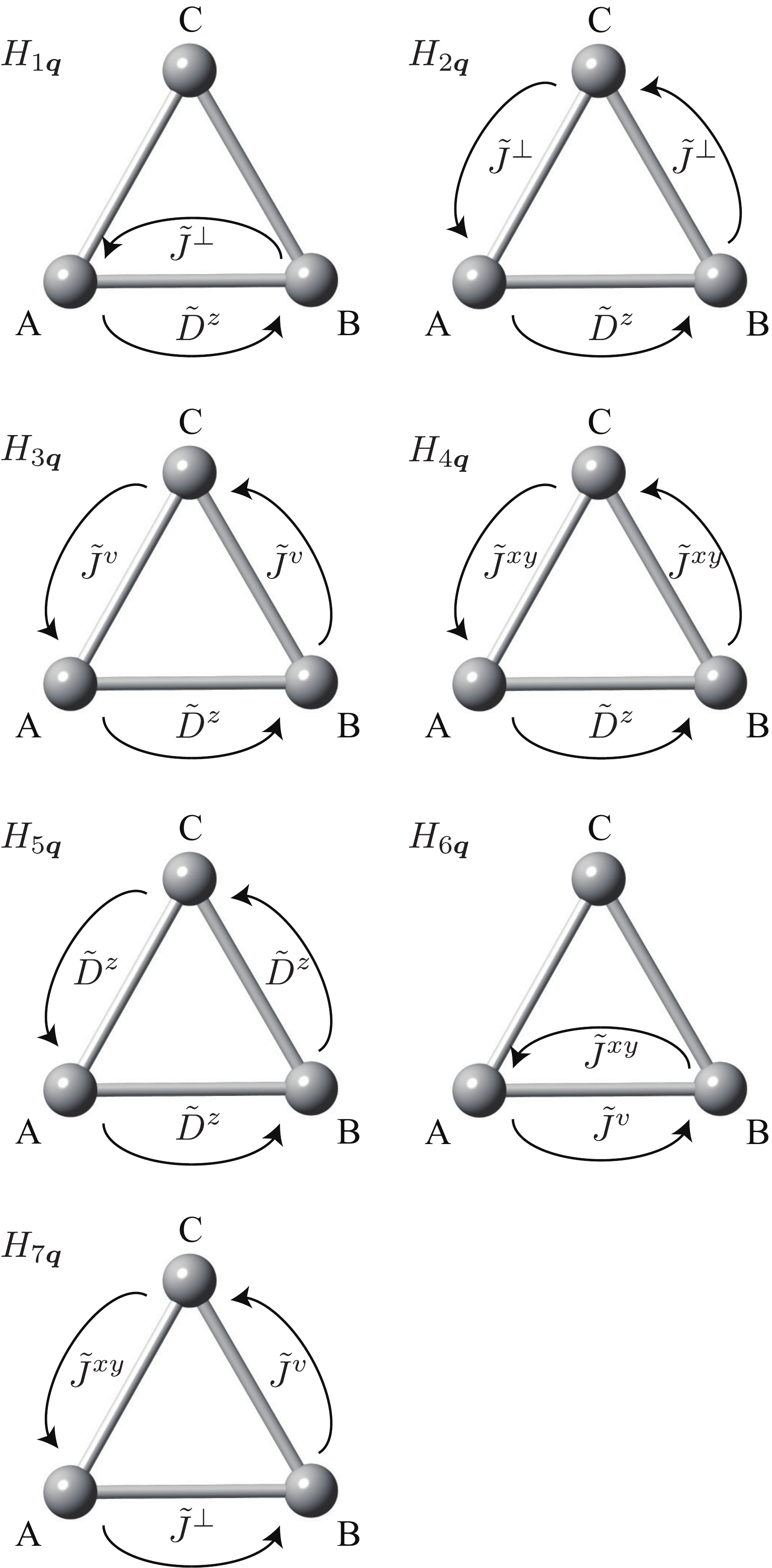} 
\caption{
\label{Fig:3sub}
Schematics of seven magnon-hopping processes giving $F^{(3)}_{\bm{q}} \neq 0$ in real space in the three-sublattice case. 
Each panel corresponds to $H_{\mu\bm{q}}$ ($\mu=1$-7) in Eq.~(\ref{eq:Fs_3sub}). 
}
\end{center}
\end{figure}

We consider a behavior of $F^{(s)}_{\bm{q}}$ in the three-sublattice case with $\eta=$ A, B, and C. 
For the general exchange interactions between different sublattices, the $3\times 3$ matrices, $\mathcal{X}_{\bm{q}}$ and $\mathcal{Y}_{\bm{q}}$, are represented by 
\begin{align}
\label{eq:HL_X_3sub}
\mathcal{X}_{\bm{q}}=&
\begin{pmatrix}
Z_{{\rm A}}&F_{{\rm AB}\bm{q}}&F_{{\rm AC}\bm{q}}\\
F^*_{{\rm AB}\bm{q}}&Z_{{\rm B}}&F_{{\rm BC}\bm{q}}\\
F^*_{{\rm AC}\bm{q}}&F^*_{{\rm BC}\bm{q}}&Z_{{\rm C}}
\end{pmatrix},\\
\label{eq:HL_Y_3sub}
\mathcal{Y}_{\bm{q}}=&
\begin{pmatrix}
 0&G_{{\rm AB}\bm{q}}&G_{{\rm AC}\bm{q}}\\
G_{{\rm AB}-\bm{q}}&0&G_{{\rm BC}\bm{q}}\\
G_{{\rm AC}-\bm{q}}&G_{{\rm BC}-\bm{q}}&0
\end{pmatrix},
\end{align}
where 
\begin{align}
\label{eq:F_3sub}
F_{\eta\eta'\bm{q}}&=\tilde{J}^{\perp} (h^{\perp{\rm (s)}}_{\eta\eta'\bm{q}}+i h^{\perp{\rm (as)}}_{\eta\eta'\bm{q}})
+i \tilde{D}^z (h^{D{\rm (s)}}_{\eta\eta'\bm{q}}+i h^{D{\rm (as)}}_{\eta\eta'\bm{q}}),\\
\label{eq:G_3sub}
G_{\eta\eta'\bm{q}}&=\tilde{J}^{v} (h^{v{\rm (s)}}_{\eta\eta'\bm{q}}+i h^{v{\rm (as)}}_{\eta\eta'\bm{q}})
+ i\tilde{J}^{xy} (h^{xy{\rm (s)}}_{\eta\eta'\bm{q}}+i h^{xy{\rm (as)}}_{\eta\eta'\bm{q}}), \\
\label{eq:Z_3sub}
Z_{{\eta}}&=J^z z_{\eta}, 
\end{align}
and $\eta,\eta'=$ A, B, and C.

The lowest contribution of $F^{(s)}_{\bm{q}}$ corresponds to the $s=3$ term similar to the two-sublattice case, which is given by 
\begin{align}
\label{eq:Fs_3sub}
F^{(3)}_{\bm{q}}=&  \tilde{D}^z \Big[\tilde{J}^{\perp} \tilde{J}^z H_{1\bm{q}} + (\tilde{J}^{\perp})^2H_{2\bm{q}}+ (\tilde{J}^{v})^2H_{3\bm{q}} \nonumber \\
 &+ (\tilde{J}^{xy})^2H_{4\bm{q}}\Big]
+(\tilde{D}^{z})^3 H_{5\bm{q}} \nonumber \\
&+ \tilde{J}^v \tilde{J}^{xy} ( \tilde{J}^z H_{6\bm{q}}+ \tilde{J}^{\perp}H_{7\bm{q}}), 
\end{align}
where $H_{\mu\bm{q}}$ ($\mu=1$-7) is the antisymmetric function consisting of odd number of $h^{\zeta{\rm (as)}}_{\bm{q}}$ and even number of $h^{\zeta{\rm (s)}}_{\bm{q}}$: $H_{\mu\bm{q}}=-H_{\mu-\bm{q}}$. 
For example, $H_{2\bm{q}}$ includes $h^{D{\rm (s)}}_{{\rm \eta\eta'}\bm{q}} h^{\perp{\rm (s)}}_{{\rm \eta'\eta''}\bm{q}} h^{\perp{\rm (as)}}_{{\rm \eta''\eta}\bm{q}}$ for $\eta \neq \eta' \neq \eta''$. 
The specific expressions of $H_{\mu\bm{q}}$ are shown in Appendix~\ref{sec:appendix} owing to the lengthy expressions.

There are mainly three contributions in the nonreciprocal magnon dispersions in Eq.~(\ref{eq:Fs_3sub}), which are proportional to $\tilde{D}^z$ including $H_{1\bm{q}}$-$H_{4\bm{q}}$, $(\tilde{D}^z)^3$ including $H_{5\bm{q}}$, and $\tilde{J}^v \tilde{J}^{xy}$ including $H_{6\bm{q}}$ and $H_{7\bm{q}}$. 
We schematically show the magnon-hopping processes corresponding to $H_{\mu\bm{q}}$ ($\mu=1$-7) in Fig.~\ref{Fig:3sub}. 
Among $H_{\mu\bm{q}}$, $H_{2\bm{q}}$, $H_{3\bm{q}}$, $H_{4\bm{q}}$, $H_{5\bm{q}}$, and $H_{7\bm{q}}$ consist of three magnon hoppings between three sublattices, while $H_{1\bm{q}}$ and $H_{6\bm{q}}$ consist of two magnon hoppings between two sublattices. 
Indeed, $H_{1\bm{q}}$ and $H_{6\bm{q}}$ correspond to the left and right panels of Fig.~\ref{Fig:2sub}, respectively, while other $H_{\mu\bm{q}}$ have no correspondence to the two-sublattice case.
In other words, this indicates that contributions from $H_{2\bm{q}}$, $H_{3\bm{q}}$, $H_{4\bm{q}}$, $H_{5\bm{q}}$, and $H_{7\bm{q}}$ can appear when the exchange interaction path includes the triangle geometry, such as the triangular and kagome lattices, while those from $H_{1\bm{q}}$ and $H_{6\bm{q}}$ do not need the triangle geometry. 
Thus, only the latter processes can contribute to the nonreciprocal magnons in the case of the one-dimensional three-sublattice chain in the absence of $F_{{\rm AC}\bm{q}}$ and $G_{{\rm AC}\bm{q}}$.

The general expression in Eq.~(\ref{eq:Fs_3sub}) describes the model parameter conditions for the nonreciprocal magnons in the three-sublattice antiferromagnetic orderings, such as the $120^{\circ}$ antiferromagnetic ordering on the triangular and breathing kagome lattices. 
We show three examples in the breathing kagome system in Secs.~\ref{sec:Nonreciprocal magnon in breathing kagome ferromagnets}, \ref{sec:Breathing kagome ferrimangets}, and \ref{sec:Breathing kagome noncollinear 120 antiferromagnets}.

\subsection{Four-sublattice case}
\label{sec:Four-sublattice case}

\begin{figure}[t!]
\begin{center}
\includegraphics[width=0.8 \hsize]{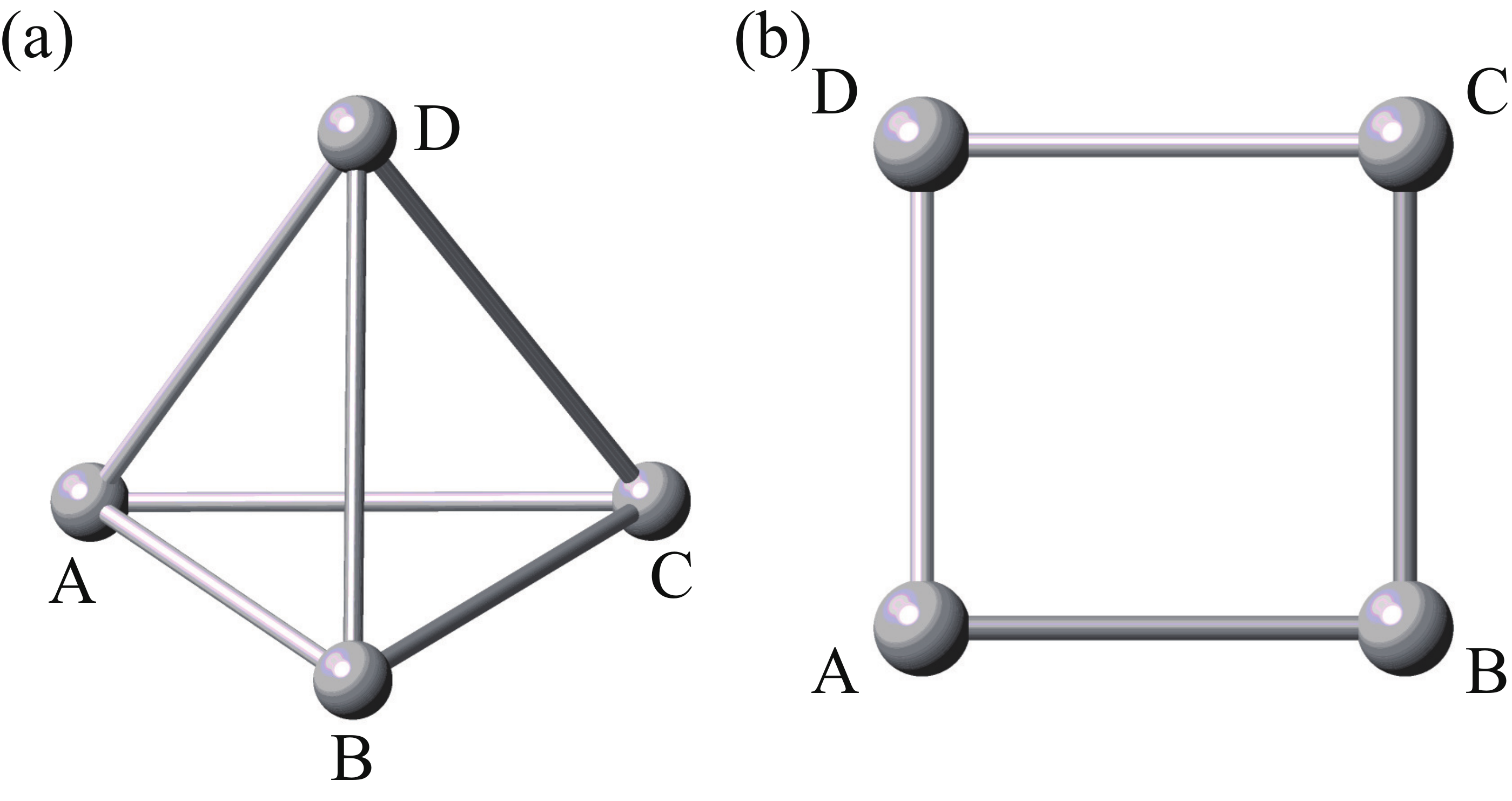} 
\caption{
\label{Fig:4sub}
Four-sublattice clusters in the shapes of (a) a tetrahedron and (b) a square. 
}
\end{center}
\end{figure}

Finally, we consider the four-sublattice case, where $\mathcal{X}_{\bm{q}}$ and $\mathcal{Y}_{\bm{q}}$ are represented by 
\begin{align}
\label{eq:HL_X_4sub}
\mathcal{X}_{\bm{q}}=&
\begin{pmatrix}
Z_{{\rm A}}&F_{{\rm AB}\bm{q}}&F_{{\rm AC}\bm{q}}&F_{{\rm AD}\bm{q}}\\
F^*_{{\rm AB}\bm{q}}&Z_{{\rm B}}&F_{{\rm BC}\bm{q}}&F_{{\rm BD}\bm{q}}\\
F^*_{{\rm AC}\bm{q}}&F^*_{{\rm BC}\bm{q}}&Z_{{\rm C}}&F_{{\rm CD}\bm{q}}\\
F^*_{{\rm AD}\bm{q}}&F^*_{{\rm BD}\bm{q}}&F^*_{{\rm CD}\bm{q}}&Z_{{\rm D}}
\end{pmatrix},\\
\label{eq:HL_Y_4sub}
\mathcal{Y}_{\bm{q}}=&
\begin{pmatrix}
 0&G_{{\rm AB}\bm{q}}&G_{{\rm AC}\bm{q}}&G_{{\rm AD}\bm{q}}\\
G_{{\rm AB}-\bm{q}}&0&G_{{\rm BC}\bm{q}}&G_{{\rm BD}\bm{q}}\\
G_{{\rm AC}-\bm{q}}&G_{{\rm BC}-\bm{q}}&0&G_{{\rm CD}\bm{q}}\\
G_{{\rm AD}-\bm{q}}&G_{{\rm BD}-\bm{q}}&G_{{\rm CD}-\bm{q}}&0
\end{pmatrix},
\end{align}
where $F_{\eta\eta'\bm{q}}$, $G_{\eta\eta'\bm{q}}$, and $Z_{{\eta}}$ are the same as Eqs.~(\ref{eq:F_3sub}), (\ref{eq:G_3sub}), and (\ref{eq:Z_3sub}), respectively.

Similar to the two- and three-sublattice cases, the lowest contribution of $F^{(s)}_{\bm{q}}$ in the four-sublattice case is $F^{(3)}_{\bm{q}}$, which is given by 
\begin{align}
\label{eq:Fs_4sub}
F^{(3)}_{\bm{q}}=&  \tilde{D}^z \Big[\tilde{J}^{\perp} \tilde{J}^z H'_{1\bm{q}} + (\tilde{J}^{\perp})^2H'_{2\bm{q}}+ (\tilde{J}^{v})^2H'_{3\bm{q}} \nonumber \\
 &+ (\tilde{J}^{xy})^2H'_{4\bm{q}}\Big]
+(\tilde{D}^{z})^3 H'_{5\bm{q}} \nonumber \\
&+ \tilde{J}^v \tilde{J}^{xy} ( \tilde{J}^z H'_{6\bm{q}}+ \tilde{J}^{\perp}H'_{7\bm{q}}), 
\end{align}
where $H'_{\mu\bm{q}}$ ($\mu=1$-7) is similar to $H_{\mu\bm{q}}$ in the three-sublattice case, and the only difference is found in the number of hopping paths due to the different number of the sublattice, as found in Appendix~\ref{sec:appendix}. 
Similar to the three-sublattice case, $H'_{2\bm{q}}$, $H'_{3\bm{q}}$, $H'_{4\bm{q}}$, $H'_{5\bm{q}}$, and $H'_{7\bm{q}}$ can appear when exchange interaction path includes the triangle geometry, while $H'_{1\bm{q}}$ and $H'_{6\bm{q}}$ do not depend on such a geometry. 
For example, in the tetrahedron cluster structure shown in Fig.~\ref{Fig:4sub}(a), all $H'_{\mu\bm{q}}$ can contribute to the nonreciprocal magnons, whereas in the square cluster structure with the nearest-neighbor exchange interactions in Fig.~\ref{Fig:4sub}(b), only $H'_{1\bm{q}}$ and $H'_{6\bm{q}}$ can contribute as
\begin{align}
\label{eq:Fs_4sub_tetra}
F^{(3)}_{\bm{q}}&=  \tilde{D}^z \tilde{J}^{\perp} \tilde{J}^z H'_{1\bm{q}} 
+ \tilde{J}^v \tilde{J}^{xy}  \tilde{J}^z H'_{6\bm{q}}. 
\end{align}
In this way, the expressions in Eqs.~(\ref{eq:Fs_4sub}) and (\ref{eq:Fs_4sub_tetra}) describe the microscopic process contributing to nonreciprocal magnons under the four-sublattice antiferromagnetic orderings, such as the pyrochlore antiferromagnets and the four-sublattice tetragonal antiferromagnets.

\section{Application to noncentrosymmetric magnets}
\label{sec:Application to noncentrosymmetric magnets}

In this section, we apply the expression in Eq.~(\ref{eq:Fs}) to noncentrosymmetric ferromagnets and antiferromagnets to host nonreciprocal magnons. 
As the ferromagnets, we consider the ferromagnetic ordering in the breathing kagome lattice structure in Sec.~\ref{sec:Nonreciprocal magnon in breathing kagome ferromagnets}. 
As the antiferromagnets, we consider three types of antiferromagnetic orderings: the staggered collinear antiferromagnetic state in the honeycomb lattice structure in Sec.~\ref{sec:Nonreciprocal magnon in honeycomb antiferromagnets}, the up-up-down ferrimagnetic state in the breathing kagome lattice structure in Sec.~\ref{sec:Breathing kagome ferrimangets}, and the noncollinear 120$^{\circ}$ antiferromagnetic state in the breathing kagome lattice structure in Sec.~\ref{sec:Breathing kagome noncollinear 120 antiferromagnets}.
In each section, we first show the Bogoliubov Hamiltonian and then we discuss magnon spectra and essential model parameters.

\subsection{Breathing kagome ferromagnets}
\label{sec:Nonreciprocal magnon in breathing kagome ferromagnets}

\subsubsection{Model}
\label{sec:Breathing kagome lattice structure}

\begin{figure}[t!]
\begin{center}
\includegraphics[width=1.0 \hsize]{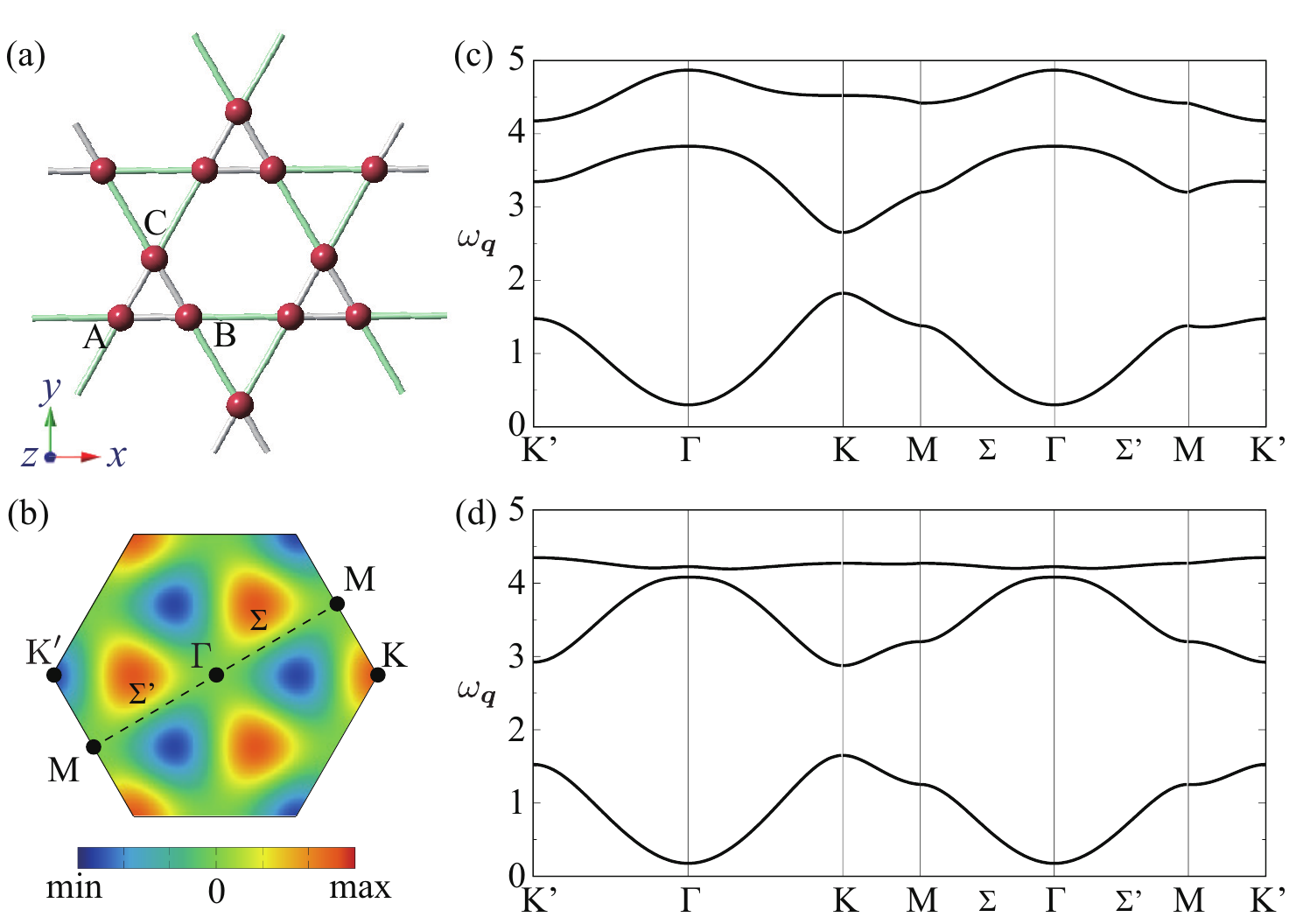} 
\caption{
\label{Fig:BKL_ferro}
(a) Breathing kagome lattice structure under the point group $D_{\rm 3h}$. 
The red spheres represent the magnetic moments along the $z$ direction. 
The different colors for bonds stand for the different magnitudes of the exchange coupling. 
(b) The first Brillouin zone in (a). 
The color plot represents angle dependence of nonreciprocal magnons characterized by $q_x (q_x^2-3q_y^2)$. 
(c, d) The magnon band structures under the ferromagnetic ordering for $D=0.2$ and $J^a=0$ (c) and $D=0$ and $J^a=0.5$ (d).
The other parameters are set as $J^{\perp}=-0.9$, $J^z=-1$, and $\gamma=0.5$. 
}
\end{center}
\end{figure}

We consider a breathing kagome lattice structure as an example of noncentrosymmetric crystal structures~\cite{matsumoto2021nonreciprocal}. 
The breathing kagome lattice structure consists of upward and downward triangles with the different sizes, as shown in Fig.~\ref{Fig:BKL_ferro}(a). 

The interaction matrix corresponding to Eq.~(\ref{eq:Jmat}) is given by 
\begin{align}
\label{eq:Jmat_BKL}
\mathcal{J}^{\triangle}_{\eta\eta'}&=
\begin{pmatrix}
J^{\perp}+J^a \cos \chi_{\eta\eta'}& D-J^a \sin \chi_{\eta\eta'}&0\\
-D-J^a \sin \chi_{\eta\eta'}& J^{\perp}-J^a \cos \chi_{\eta\eta'}&0\\
0&0&J^z
\end{pmatrix}, \\
\label{eq:Jmat_BKL2}
\mathcal{J}^{\bigtriangledown}_{\eta\eta'}&=\gamma\mathcal{J}^{\triangle}_{\eta\eta'}, 
\end{align}
where the superscript $\triangle$ ($\bigtriangledown$) denotes the interaction for the upward (downward) triangles where $\gamma$ is the breathing parameter, and $\chi_{{\rm AB}}=0$, $\chi_{{\rm BC}}=2\pi/3$ and $\chi_{{\rm CA}}=4\pi/3$. 
We here consider four independent interactions from the symmetry analysis: the isotropic inplane interaction $J^{\perp}$, the DM interaction $D$, the bond-dependent anisotropic interaction $J^a$, and the $z$ spin interaction $J^z$. 
The direction of the DM vector is taken along the $+z$ ($-z$) direction for the upward (downward) triangle.
The anisotropic interactions, $D$, $J^a$, and $J^z-J^{\perp}$ originates from the relativistic spin-orbit coupling and/or dipole-diople interactions. 
Compared to Eq.~(\ref{eq:Jmat}), one finds the correspondence of $(J^{v}_{\eta\eta'}, J^{xy}_{\eta\eta'})$ and $(J^a \cos \chi_{\eta\eta'}, -J^a \sin \chi_{\eta\eta'})$. 

In the ferromagnetic state with magnetic moments along the $z$ direction, we do not need the rotation of the spin frame, i.e., 
$\tilde{J}^{\perp}_{\eta\eta'}=J^{\perp}$, $\tilde{D}_{\eta\eta'}=D$, $\tilde{J}^v_{\eta\eta'}=J^a \cos \chi_{\eta\eta'}$, $\tilde{J}^{xy}_{\eta\eta'}=-J^a \sin \chi_{\eta\eta'}$, and $\tilde{J}_{\eta\eta'}^{z}=J^z$ in Eqs.~(\ref{eq:rot_J})-(\ref{eq:rot_z}). 
By performing the Holstein-Primakov transformation and then the Fourier transformation, the $3\times 3$ matrices $\mathcal{X}_{\bm{q}}$ and $\mathcal{Y}_{\bm{q}}$ in the Bogoliubov Hamiltonian matrix $H^{\rm B}_{\bm{q}}$ are given by~\cite{matsumoto2021nonreciprocal} 
\begin{align}
\label{eq:Xq_BKL}
\mathcal{X}_{\bm{q}}=&
\begin{pmatrix}
Z&F_{{\rm AB}{\bm{q}}}&F^{*}_{{\rm CA}{\bm{q}}}\\
F^{*}_{{\rm AB}{\bm{q}}}&Z&F_{{\rm BC}{\bm{q}}}\\
F_{{\rm CA}{\bm{q}}}&F^{*}_{{\rm BC}{\bm{q}}}&Z
\end{pmatrix},\\
\label{eq:Yq_BKL}
\mathcal{Y}_{\bm{q}}=&
\begin{pmatrix}
0&G_{{\rm AB}{\bm{q}}}&G_{{\rm CA}{-\bm{q}}}\\
G_{{\rm AB}-{\bm{q}}}&0&G_{{\rm BC}{\bm{q}}}\\
G_{{\rm CA}{\bm{q}}}&G_{{\rm BC}{-\bm{q}}}&0
\end{pmatrix}
,\end{align}
where 
\begin{align}
\label{eq:F_BKL}
F_{\eta\eta'{\bm{q}}}
=&
\left(J^{\perp}-iD\right)
\left(e^{i \bm{q} \cdot \bm{\rho}_{\eta\eta'}}
+\gamma
e^{-i \bm{q} \cdot \bm{\rho}_{\eta\eta'}}\right),\\
\label{eq:G_BKL}
G_{\eta\eta'{\bm{q}}}
=&
J^a e^{-i \chi_{\eta\eta'}}
\left(e^{i \bm{q} \cdot \bm{\rho}_{\eta\eta'}}
+\gamma  
e^{-i \bm{q} \cdot \bm{\rho}_{\eta\eta'}}\right),\\
Z=&-2(1+ \gamma )J^{z},
\end{align}
where $\bm{\rho}_{\eta\eta'}$ is the displacement vector between $\eta$ and $\eta'$ sublattices in the breathing kagome lattice structure. 
It is noted that the length of a side of both the upward and downward triangles is taken as one for notational simplicity. 

\subsubsection{Result}
\label{Result_BKL_FM}
 
The ferromagnetic spin configuration becomes stable when $J^z$ is dominant and ferromagnetic. 
We show the magnon dispersions along high symmetry lines in the Brillouin zone [Fig.~\ref{Fig:BKL_ferro}(b)] in the ferromagnetic state after the numerical Bogoliubov transformation. 
Figure~\ref{Fig:BKL_ferro}(c) shows the magnon spectra $\omega_{\bm{q}}$ for $D=0.2$ without $J^a$, while Fig.~\ref{Fig:BKL_ferro}(d) shows ones for $J^a=0.5$ without $D$. 
Both cases clearly exhibit that the magnon bands are modulated antisymmetrically in the functional form of $q_x (q_x^2-3q_y^2)$~\cite{matsumoto2021nonreciprocal}. 
The angle dependence in the limit of $|\bm{q}| \to 0$ is given by $\cos 3\phi$ when setting $(q_x, q_y)=q (\cos \phi, \sin \phi)$, as shown in Fig.~\ref{Fig:BKL_ferro}; the antisymmetric modulation appears along the K'-$\Gamma$-K line, while it does not along the M($\Sigma$)-$\Gamma$-M($\Sigma'$) line.  

The above result means that both $D$ and $J^a$ become the origin of the nonreciprocal magnons. 
Such model parameter conditions are easily obtained by evaluating $F^{(s)}_{\bm{q}}$ in Eq.~(\ref{eq:Fs}) without solving the eigenvalue problems. 
For a general case at $D \neq 0$ and $J^a \neq 0$, the lowest-order contribution from $F^{(s)}_{\bm{q}}$ is of third order as shown in Sec.~\ref{sec:Three-sublattice case}, which is given by  
\begin{align}
\label{eq:F_BKL_FM_general}
F^{(3)}_{\bm{q}}=&-12  \gamma (1-\gamma)(\sqrt{3} J^{\perp}+D) \nonumber \\
&\times [2 D (\sqrt{3} J^{\perp}-D)+3 (J^{a})^2 ]  f^{3\phi}_{\bm{q}},
\end{align}
where 
  \begin{align}
  f^{3\phi}_{\bm{q}}= \left(\cos q_x-\cos \sqrt{3} q_y \right) \sin q_x. 
\end{align}
Thus, one finds that the antisymmetric functional form of $f^{3\phi}_{\bm{q}}= (\cos q_x-\cos \sqrt{3} q_y) \sin q_x$ in $F^{(3)}_{\bm{q}}$ is consistent with that in the magnon dispersions in Figs.~\ref{Fig:BKL_ferro}(c) and \ref{Fig:BKL_ferro}(d). 
Furthermore, the expression in Eq.~(\ref{eq:F_BKL_FM_general}) clearly presents the essential parameters in nonreciprocal magnons: $\gamma$, $D$, and $J^{a}$. 
The condition of $\gamma \neq 1$ represents the importance of the breathing structure, which is reasonable in terms of spatial inversion symmetry; it is recovered for $\gamma=1$. 
In a similar way, $F^{(3)}_{\bm{q}}$ shows that no antisymmetric magnon dispersions appear when $D=-\sqrt{3} J^{\perp} $. 
This is rather surprising, as such a condition is not obtained by the symmetry argument. 
Indeed, we confirmed that the magnon dispersions become symmetric at $D=-\sqrt{3} J^{\perp} $. 

The other essential parameters are $D$ and $J^a$, as inferred from the results in Figs.~\ref{Fig:BKL_ferro}(c) and \ref{Fig:BKL_ferro}(d). 
In the case of Fig.~\ref{Fig:BKL_ferro}(c) for nonzero $D$ and $J^a=0$, Eq.~(\ref{eq:F_BKL_FM_general}) reduces to  
 \begin{align}
 \label{eq:F_BKL_FM_D}
F^{(3)}_{\bm{q}}= -24 \gamma (1-\gamma) D (3  J^{\perp 2}-  D^2) f^{3\phi}_{\bm{q}}. 
 \end{align}
The result indicates that asymmetric feature vanishes for $D =0$ and $D=\sqrt{3} J^{\perp}$ in addition to $\gamma \neq 0, 1$ and $D=-\sqrt{3} J^{\perp} $ in Eq.~(\ref{eq:F_BKL_FM_general}). 
Thus, $D$ is one of the essential parameters, and its odd order contributes to the asymmetric dispersions. 
On the other hand, for nonzero $J^a$ and $D=0$, Eq.~(\ref{eq:F_BKL_FM_general}) turns into 
 \begin{align}
  \label{eq:F_BKL_FM_Ja}
F^{(3)}_{\bm{q}}= -36 \sqrt{3}  \gamma (1-\gamma) J^{\perp} (J^a)^2 f^{3\phi}_{\bm{q}}. 
 \end{align}
We find that the even order of $J^a$ becomes the essential parameters in the case of $D=0$. 
These results are consistent with those obtained from the general expression in Sec.~\ref{sec:Three-sublattice case}.

\subsection{Honeycomb antiferromagnets}
\label{sec:Nonreciprocal magnon in honeycomb antiferromagnets}

\subsubsection{Model}
\label{sec:Honeycomb lattice structure}

\begin{figure}[t!]
\begin{center}
\includegraphics[width=1.0 \hsize]{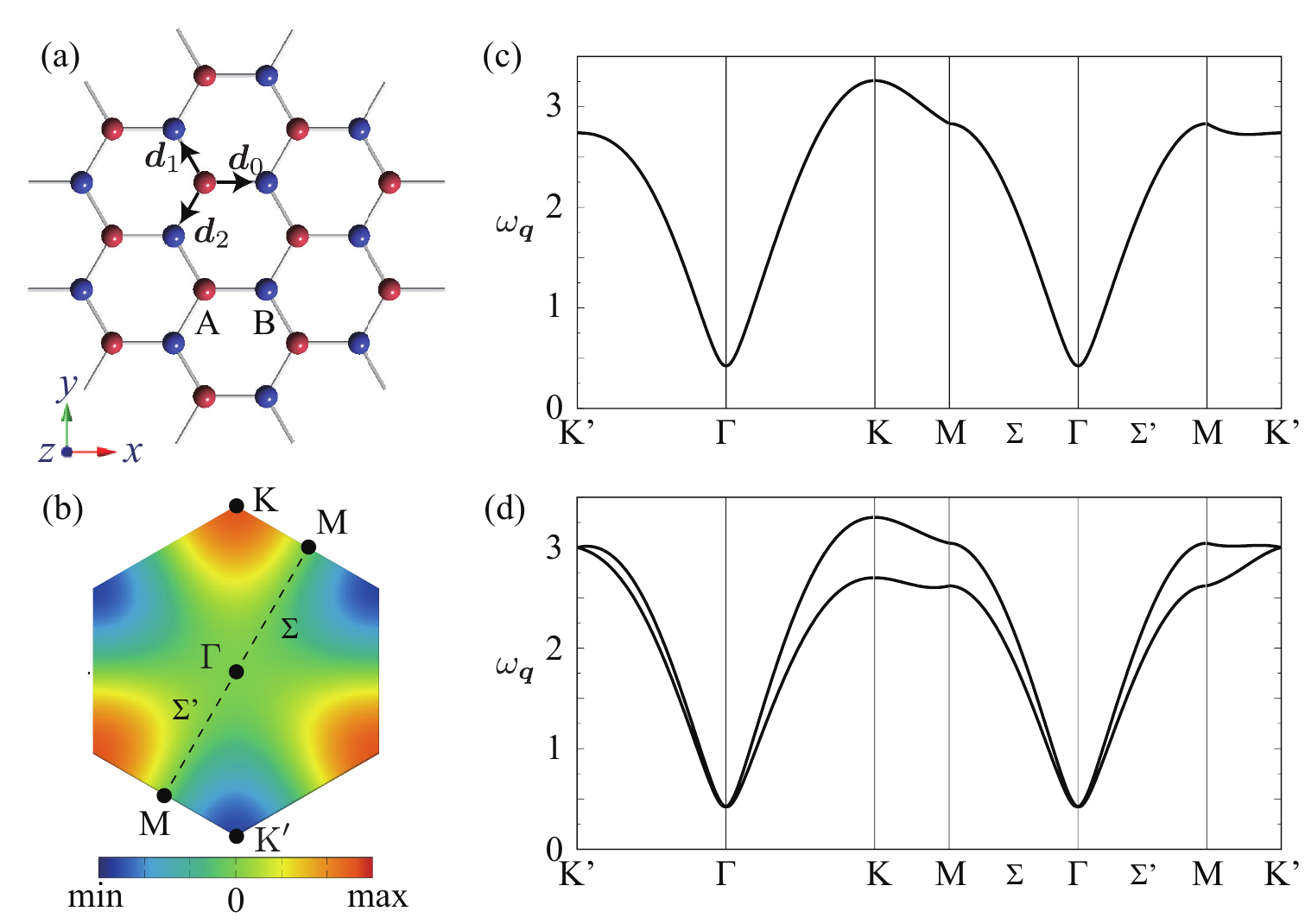} 
\caption{
\label{Fig:HL_AFM}
Honeycomb lattice structure under the point group $D_{\rm 6h}$. 
The red (blue) spheres represent the up (down) spins along the $z$ direction. 
The three bond vectors, $\bm{d}_0$, $\bm{d}_1$, and $\bm{d}_2$, are also shown.
(b) The first Brillouin zone in (a). 
The color plot represents angle dependence of nonreciprocal magnons characterized by $q_y (q_y^2-3q_x^2)$. 
(c, d) The magnon band structures under the staggered antiferromagnetic ordering for $D=0.05$ and $J^a=0$ (c) and $D=0$ and $J^a=0.1$ (d).
The other parameters are set as $J^{\perp}=0.99$ and $J^z=1$. 
}
\end{center}
\end{figure}

The honeycomb lattice structure consists of two sublattices A and B, as shown in Fig.~\ref{Fig:HL_AFM}(a). 
From the presence of threefold rotational symmetry around the $z$ axis and mirror symmetry perpendicular to the $xy$ plane along the bond direction at each local site, the interaction tensor for the nearest-neighbor spins is given by
\begin{align}
\label{eq:Jmat_HL1}
\mathcal{J}_{\rm AB}^{\nu}&=
\begin{pmatrix}
J^{\perp}+J^a \cos \chi_{\nu}& -J^a \sin \chi_{\nu}&0\\
-J^a \sin \chi_{\nu}& J^{\perp}-J^a \cos \chi_{\nu}&0\\
0&0&J^z
\end{pmatrix},  
\end{align}
where $\nu=0$-$2$ is the bond index for the nearest-neighbor spins and $\chi_{\nu}=0, 2\pi/3, 4\pi/3$ for $\nu=0$-$2$. 
The three bond vectors are $\bm{d}_0=(1,0)$, $\bm{d}_1=(-1/2,\sqrt{3}/2)$, and $\bm{d}_2=(-1/2,-\sqrt{3}/2)$. 
The DM interaction vanishes owing to inversion symmetry on the A-B bond center. 
The contribution of the DM interaction arises in the interaction tensor for the next-nearest-neighbor spins belonging to the same sublattice, which is given by 
\begin{align}
\mathcal{J}^{\nu'}_{\rm AA}&=-\mathcal{J}^{\nu'}_{\rm BB}=
\begin{pmatrix}
0 & D&0\\
-D & 0&0\\
0&0&0
\end{pmatrix}, 
\end{align}
where $\nu'=0$-$5$ is the bond index for the next-nearest-neighbor spins. 
We ignore the other symmetric exchange interactions in $\mathcal{J}_{\rm AA}$ and $\mathcal{J}_{\rm BB}$. 
The opposite sign of the DM interaction for the A and B sublattices is owing to  inversion symmetry in the system. 

We consider the staggered antiferromagnetic state with $S^z_{\rm A}=1$ and $S^z_{\rm B}=-1$, as schematically shown in Fig.~\ref{Fig:HL_AFM}(a). 
In contrast to the ferromagnetic ordering in Sec.~\ref{sec:Nonreciprocal magnon in breathing kagome ferromagnets}, the spin frame is required to be locally rotated according to Eq.~(\ref{eq:spinrotate}) in order to use Eq.~(\ref{eq:Fs}). 
After rotating the spin frame, the effective interactions corresponding to Eqs.~(\ref{eq:rot_J})-(\ref{eq:rot_z})  are given by 
\begin{align}
\label{eq:Jtilde_HL_Jperp}
\tilde{J}_{\rm AB}^{\perp (\nu)}&=-J^a \cos \chi_{\nu}, \\
\label{eq:Jtilde_HL_Jv}
\tilde{J}_{\rm AB}^{v}&=-J^{\perp}, \\
\label{eq:Jtilde_HL_Jz}
\tilde{J}_{\rm AB}^z&=-J^z, \\
\label{eq:Jtilde_HL_DAB}
\tilde{D}^{(\nu)}_{\rm AB}&=-J^a  \sin \chi_{\nu}, \\
\label{eq:Jtilde_HL_DAA}
\tilde{D}_{\rm AA}&=\tilde{D}_{\rm BB}=D, 
\end{align}
for the $\nu$th bond ($\tilde{J}_{\rm AB}^{v}$ and $\tilde{J}_{\rm AB}^{z}$ do not depend on $\nu$). 
Owing to the $\pi$ rotation of the spin frame for the sublattice B, the bond-dependent interaction $J^a$ is transformed into $\tilde{J}_{\rm AB}^{\perp}$ and $\tilde{D}_{\rm AB}$ in Eqs.~(\ref{eq:Jtilde_HL_Jperp}) and (\ref{eq:Jtilde_HL_DAB}), and the sublattice-dependent DM interaction turns into the uniform DM interaction in Eq.~(\ref{eq:Jtilde_HL_DAA}). 
By performing the Holstein-Primakov transformation and then the Fourier transformation, the $2\times 2$ matrices $\mathcal{X}_{\bm{q}}$ and $\mathcal{Y}_{\bm{q}}$ in Eq.~(\ref{eq:Ham_B2}) are given by~\cite{Hayami_doi:10.7566/JPSJ.85.053705,Matsumoto_PhysRevB.101.224419} 
\begin{align}
\label{eq:HL_X}
\mathcal{X}_{\bm{q}}=&
\begin{pmatrix}
Z_{\bm{q}}&F_{\bm{q}}\\
F^*_{\bm{q}}&Z_{\bm{q}}
\end{pmatrix},\\
\label{eq:HL_Y}
\mathcal{Y}_{\bm{q}}=&
\begin{pmatrix}
 0&G_{\bm{q}}\\
G_{-\bm{q}}&0
\end{pmatrix},
\end{align}
where 
\begin{align}
F_{\bm{q}}&=-J^a \sum_{\nu}   e^{i(\bm{q} \cdot \bm{d}_{\nu}-\chi_{\nu}) },\\
G_{\bm{q}}&=-J^{\perp} \sum_{\nu}   e^{i \bm{q} \cdot \bm{d}_{\nu} }, \\
Z_{\bm{q}}&=3J^z+4D   \left( \cos \frac{3q_x}{2} -\cos \frac{\sqrt{3}q_y}{2}  \right)\sin \frac{\sqrt{3}q_y}{2}. 
\end{align}

\subsubsection{Result}

The staggered antiferromagnetic spin configuration is stabilized by supposing that $J^z$ is the dominant antiferromagnetic interaction. 
We take $J^z=1$ and $J^{\perp}=0.99$, respectively. 
The magnon dispersions in the antiferromagnetic state are shown in Figs.~\ref{Fig:HL_AFM}(c) and \ref{Fig:HL_AFM}(d), where the Brillouin zone is shown in Fig.~\ref{Fig:HL_AFM}(b). 
The magnon spectra $\omega_{\bm{q}}$ in Fig.~\ref{Fig:HL_AFM}(c) are calculated for $D=0.05$ and $J^a=0$ and those in Fig.~\ref{Fig:HL_AFM}(d) are for $D=0$ and $J^a=0.1$. 
Similar to the result in Sec.~\ref{sec:Nonreciprocal magnon in breathing kagome ferromagnets}, the asymmetric modulations occur in both situations. 
The antisymmetric functional form is given by $q_y (3q_x^2-q_y^2)$, as shown by the color plot in Fig.~\ref{Fig:HL_AFM}(b), which means that the angle dependence is expressed as $\sin 3\phi$ in the limit of $|\bm{q}| \to 0$. 

From Eq.~(\ref{eq:Fs}), the essential model parameters are straightforwardly computed. 
The lowest-order contribution in terms of $D$ is given by 
\begin{align}
\label{eq:F1_HC}
F^{(1)}_{\bm{q}}
=& 8 (\tilde{D}_{\rm AA}+\tilde{D}_{\rm BB}) \left(\cos \frac{3 q_x}{2}-\cos \frac{\sqrt{3} q_y}{2}\right) \sin \frac{\sqrt{3} q_y}{2}  \\
=&16 D \left(\cos \frac{3 q_x}{2}-\cos \frac{\sqrt{3} q_y}{2}\right) \sin \frac{\sqrt{3} q_y}{2}. 
\end{align}
Meanwhile, the lowest-order contribution in terms of $J^a$ is of third-order, which is given by
\begin{align}
\label{eq:F3_HC}
F^{(3)}_{\bm{q}}=&72 \tilde{J}^z_{\rm AB} 
\bigg[\sin \sqrt{3} q_y (\tilde{D}^{(2)}_{\rm AB} \tilde{J}_{\rm AB}^{\perp (1)}-\tilde{D}^{(1)}_{\rm AB} \tilde{J}_{\rm AB}^{\perp (2)}) \nonumber \\
&-\tilde{J}_{\rm AB}^{\perp (0)} \bigg\{ 
\tilde{D}^{(1)}_{\rm AB} \sin \left( \frac{3 q_x+\sqrt{3} q_y}{2} \right) \nonumber \\
&+\tilde{D}^{(2)}_{\rm AB} \sin \left(\frac{3 q_x-\sqrt{3} q_y}{2} \right)\bigg\}\bigg] \\
=&72 \sqrt{3} J^z(J^a)^2 \left(\cos \frac{3 q_x}{2}-\cos \frac{\sqrt{3} q_y}{2}\right) \sin \frac{\sqrt{3} q_y}{2}, 
\end{align}
where we set $D=0$. 
These results are consistent with those in Eqs.~(\ref{eq:Fs_1sub_D}) and (\ref{eq:Fs_2sub}) in Sec.~\ref{sec:General feature of F}.
Similar to the ferromagnetic ordering in Sec.~\ref{sec:Nonreciprocal magnon in breathing kagome ferromagnets}, the result obtained from Eq.~(\ref{eq:Fs}) gives the same functional form as that in the magnon dispersions in Figs.~\ref{Fig:HL_AFM}(c) and \ref{Fig:HL_AFM}(d). 
Furthermore, the expressions in Eqs.~(\ref{eq:F1_HC}) and (\ref{eq:F3_HC}) indicate the odd order of the effective DM interaction causes the asymmetric magnon dispersions as obtained in Sec.~\ref{sec:General feature of F}.

\subsection{Breathing kagome ferrimangets}
\label{sec:Breathing kagome ferrimangets}

\subsubsection{Model}

\begin{figure}[t!]
\begin{center}
\includegraphics[width=1.0 \hsize]{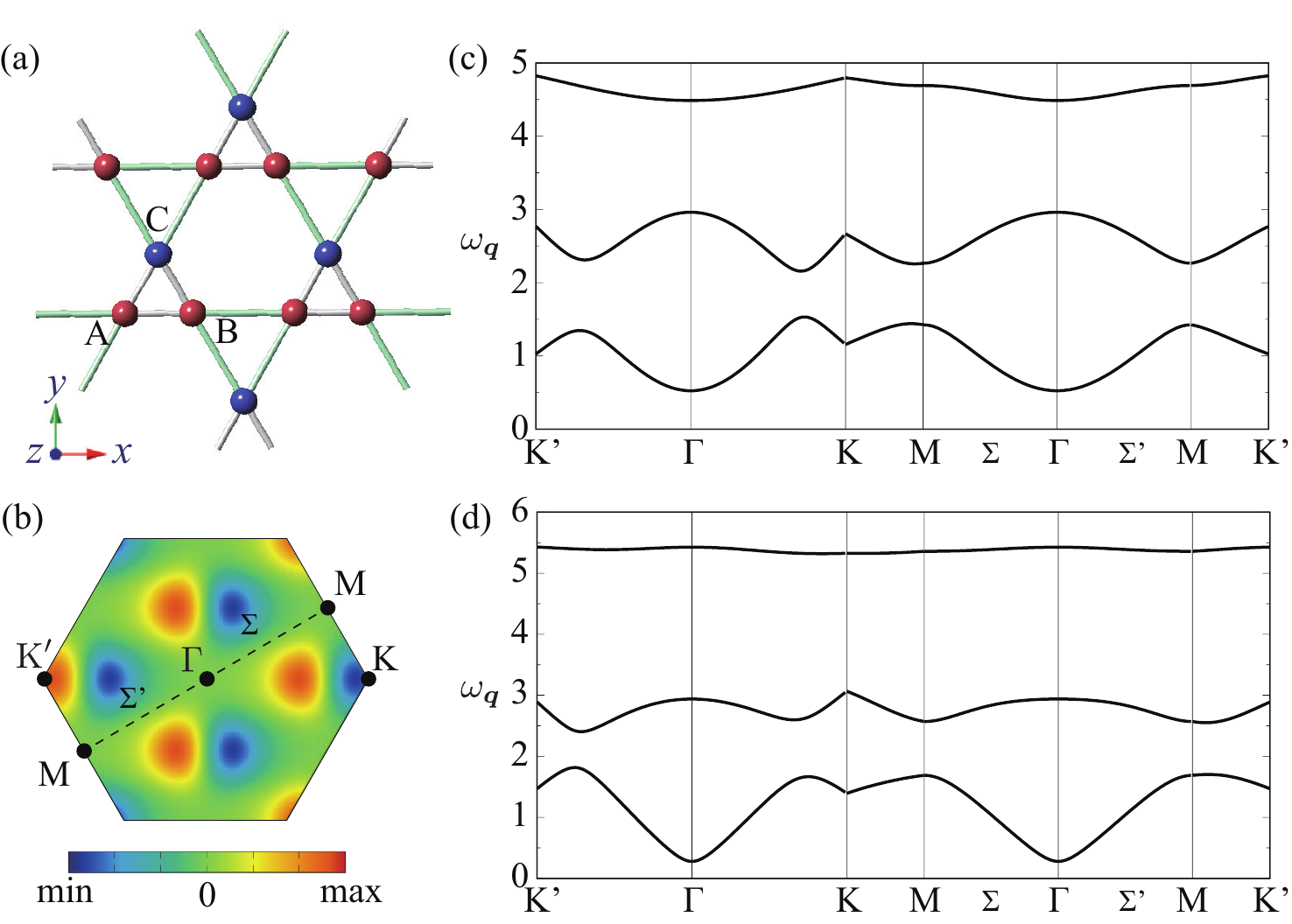} 
\caption{
\label{Fig:BKL_uud}
Breathing kagome lattice structure under the point group $D_{\rm 3h}$. 
The red (blue) spheres represent the up (down) spins along the $z$ direction. 
(b) The first Brillouin zone in (a). 
The color plot represents angle dependence of nonreciprocal magnons characterized by a linear combination of $q_x (q_x^2-3q_y^2)$ and $q_x(q_x^2-q_y^2) (q_x^2-3q_y^2)$. 
(c, d) The magnon band structures under the up-up-down magnetic ordering for $D=0.2$, $J^a=0$, and $J^{\parallel}=-2$ (c) and $D=0$, $J^a=0.5$, and $J^{\parallel}=-2.4$ (d).
The other parameters are set as $J^{\perp}=0.9$, $J^z=1$, and $\gamma=0.5$. 
}
\end{center}
\end{figure}

We discuss the other example of the nonreciprocal magnons in the ferrimagnetic state. 
We consider the up-up-down magnetic ordering in the breathing kagome lattice structure as a fundamental example. 
The up-up-down spin configuration is shown in Fig.~\ref{Fig:BKL_uud}(a). 

The spin Hamiltonian is common to Eqs.~(\ref{eq:Jmat_BKL}) and (\ref{eq:Jmat_BKL2}) in Sec.~\ref{sec:Nonreciprocal magnon in breathing kagome ferromagnets}. 
The effective interaction tensors corresponding to Eqs.~(\ref{eq:rot_J})-(\ref{eq:rot_z}) are modified from those in Sec.~\ref{sec:Nonreciprocal magnon in breathing kagome ferromagnets} for the antiparallel spin pairs, i.e., A-C and B-C spins. 
The interactions are given by
\begin{align}
\tilde{J'}_{\rm CA}^{\perp}=&-J^a \cos{\chi_{\rm CA}},\\
\tilde{J'}_{\rm BC}^{\perp}=&-J^a \cos{\chi_{\rm BC}},\\
\tilde{J'}_{\rm CA}^{v}=&\tilde{J'}_{\rm BC}^{v}
=-J^{\perp} ,\\
\tilde{J'}_{\rm CA}^{z}=&\tilde{J'}_{\rm BC}^{z}
=-J^z ,\\
\tilde{J'}_{\rm CA}^{xy}
=&-D ,\\
\tilde{J'}_{\rm BC}^{xy}
=&D ,\\
\tilde{D'}_{\rm CA}
=& J^a \sin \chi_{\rm CA}, \\
\tilde{D'}_{\rm BC}
=& -J^a \sin \chi_{\rm BC}. 
\end{align}
The $\pi$ rotation of the spin frame around the $y$ axis for the C sublattice leads to the correspondence between ($\tilde{J}'^{\perp}_{\eta\eta'}, \tilde{D}'_{\eta\eta'} \leftrightarrow J'^{v}_{\eta\eta'}, J'^{xy}_{\eta\eta'}$) and ($\tilde{J}'^{v}_{\eta\eta'}, \tilde{J}'^{xy}_{\eta\eta'} \leftrightarrow J'^{\perp}_{\eta\eta'}, D'_{\eta\eta'}$). 

Then, the $3\times3$ matrices $\mathcal{X}_{\bm{q}}$ and $\mathcal{Y}_{\bm{q}}$ in the Bogoliubov Hamiltonian in momentum space are obtained as~\cite{matsumoto2021nonreciprocal}  
\begin{align}
\label{eq:Xq_BKL_uud}
\mathcal{X}_{\bm{q}}=&
\begin{pmatrix}
0&F_{{\rm AB}{\bm{q}}}&F'^{*}_{{\rm CA}{\bm{q}}}\\
F^{*}_{{\rm AB}{\bm{q}}}&0&F'_{{\rm BC}{\bm{q}}}\\
F'_{{\rm CA}{\bm{q}}}&F'^{*}_{{\rm BC}{\bm{q}}}&Z
\end{pmatrix},\\
\label{eq:Yq_BKL_uud}
\mathcal{Y}_{\bm{q}}=&
\begin{pmatrix}
0&G_{{\rm AB}{\bm{q}}}&G'_{{\rm CA}{-\bm{q}}}\\
G_{{\rm AB}-{\bm{q}}}&0&G'_{{\rm BC}{\bm{q}}}\\
G'_{{\rm CA}{\bm{q}}}&G'_{{\rm BC}{-\bm{q}}}&0
\end{pmatrix},
\end{align}
where 
\begin{align}
F'_{{\rm BC}{\bm{q}}}
=&
-J^a e^{-i \chi_{{\rm BC}}}
\left(e^{i \bm{q} \cdot \bm{\rho}_{{\rm BC}}}
+\gamma 
e^{-i \bm{q} \cdot \bm{\rho}_{{\rm BC}}}\right),\\
F'_{{\rm CA}{\bm{q}}}
=&
-J^a e^{i \chi_{\rm CA}}
\left(e^{i \bm{q} \cdot \bm{\rho}_{\rm CA}}
+\gamma 
e^{-i \bm{q} \cdot \bm{\rho}_{\rm CA}}\right),\\
G'_{{\rm BC}{\bm{q}}}
=&
\left(-J^{\perp}+i D\right)
\left(e^{i \bm{q} \cdot \bm{\rho}_{{\rm BC}}}
+\gamma 
e^{-i \bm{q} \cdot \bm{\rho}_{{\rm BC}}}\right),\\
G'_{{\rm CA}{\bm{q}}}
=&
\left(-J^{\perp}-iD\right)
\left(e^{i \bm{q} \cdot \bm{\rho}_{\rm CA}}
+\gamma
e^{-i \bm{q} \cdot \bm{\rho}_{\rm CA}}\right),\\
Z=&2(1+ \gamma )J^{z}. 
\end{align}
$F_{{\rm AB}{\bm{q}}}$ and $G_{{\rm AB}{\bm{q}}}$ are common to Eqs.~(\ref{eq:F_BKL}) and (\ref{eq:G_BKL}), respectively.

\subsubsection{Result}

The up-up-down spin configuration is not simply stabilized by the spin Hamiltonian owing to the degeneracy arising from the kagome lattice structure.
We here introduce the interlayer ferromagnetic exchange coupling with the coupling constant $J^{\parallel}$ by supposing the quasi-two-dimensional structure~\cite{matsumoto2021nonreciprocal}. 
Then, the diagonal matrix element $(\mathcal{X}_{\bm{q}})_{ii}=(0,0,Z)$ in Eq.~(\ref{eq:Xq_BKL_uud}) turns into $(\mathcal{X}_{\bm{q}})_{ii}=(J^{\parallel},J^{\parallel},Z+J^{\parallel})$, which opens the gap in the magnon spectra.  
In the following, we fix $J^{\perp}=0.9$, $J^z=1$, and $\gamma=0.5$. 

Figures~\ref{Fig:BKL_uud}(c) and \ref{Fig:BKL_uud}(d) show the magnon dispersions under the up-up-down magnetic ordering along high symmetry lines in the Brillouin zone in Fig.~\ref{Fig:BKL_uud}(b). 
The data in Fig.~\ref{Fig:BKL_uud}(c) is obtained at $D=0.2$, $J^a=0$, and $J^{\parallel}=-2$ and that in Fig.~\ref{Fig:BKL_uud}(d) is $D=0$, $J^a=0.5$, and $J^{\parallel}=-2.4$.
In contrast to the magnon dispersions in the ferromagnetic state in Sec.~\ref{sec:Nonreciprocal magnon in breathing kagome ferromagnets}, threefold rotational symmetry in the dispersions does not hold, which is consistent with the symmetry of the magnetic orderings. 
This result indicates that there is an additional angle dependence of $\cos \phi$ to $\cos 3\phi$, whose behavior is schematically shown as the color plot in Fig.~\ref{Fig:BKL_uud}(b).  
We also confirm that the magnon dispersions in Figs.~\ref{Fig:BKL_uud}(c) and \ref{Fig:BKL_uud}(d) are characterized by the above angle dependence. 

By evaluating $F^{(s)}_{\bm{q}}$ in Eq.~(\ref{eq:Fs}), the essential model parameters are extracted. 
The lowest-order contribution is given as the same form of Eq.~(\ref{eq:F_BKL_FM_general}) except for the sign. 
In other words, the lowest-order contribution gives the angle dependence of $\cos 3 \phi$. 
The other $\cos \phi$ dependence is obtained by the second lowest-order contribution $F^{(5)}_{\bm{q}}$. 
For $J^a=0$, $F^{(5)}_{\bm{q}}$ is given by 
\begin{align}
\label{eq:F5_kagome_uud_D}
F^{(5)}_{\bm{q}}
=&10 \gamma^2(1-\gamma) h_1  \Big[\tilde{D}_{\rm AB} (\tilde{J'}_{\rm BC}^{v}\tilde{J'}_{\rm CA}^{v}+\tilde{J'}_{\rm BC}^{xy} \tilde{J'}_{\rm CA}^{xy}) \nonumber \\ 
&+\tilde{J}_{\rm AB} (\tilde{J'}_{\rm BC}^{v} \tilde{J'}_{\rm CA}^{xy}- \tilde{J'}_{\rm CA}^{v} \tilde{J'}_{\rm BC}^{xy}) \Big] q^5 \cos (a)
\\
=&40  \gamma^2 (1-\gamma)  D (3 J^{\perp 2}-D^2) (J^{\perp 2}+D^2) q^5 \cos \phi, 
\end{align}
where $h_1=2 \tilde{D}_{\rm AB}^2+2 \tilde{J}_{\rm AB}^2+(\tilde{J'}_{\rm BC}^{v})^2+(\tilde{J'}_{\rm CA}^{v})^2+(\tilde{J'}_{\rm BC}^{xy})^2+(\tilde{J'}_{\rm CA}^{xy})^2$. 
On the other hand, for $D=0$, $F^{(5)}_{\bm{q}}$ is represented by 
\begin{align}
\label{eq:F5_kagome_uud_Ja}
F^{(5)}_{\bm{q}}
=&10 \gamma^2(1-\gamma)   h_2 \Big[\tilde{D'}_{\rm BC} (\tilde{J}_{\rm AB}^{\perp} \tilde{J'}_{\rm CA}^{\perp}- \tilde{J}_{\rm AB}^{v} \tilde{J'}_{\rm CA}^{v}) \nonumber \\
&+\tilde{D'}_{\rm CA} (\tilde{J}_{\rm AB}^{\perp} \tilde{J'}_{\rm BC}^{\perp}- \tilde{J}_{\rm AB}^{v} \tilde{J'}_{\rm BC}^{v})\Big] q^5 \cos \phi
\\
=&60 \sqrt{3} \gamma^2 (1-\gamma)  J^{\perp} (J^a)^2  [J^{\perp 2}-(J^a)^2] q^5 \cos \phi, 
\end{align}
where we omit the irrelevant contributions and $h_2=\tilde{D'}_{\rm BC}^2+\tilde{D'}_{\rm CA}^2-2 (\tilde{J}_{\rm AB}^{\perp})^2+(\tilde{J'}_{\rm BC}^{\perp})^2+(\tilde{J'}_{\rm CA}^{\perp})^2+2 (\tilde{J}_{\rm AB}^{v})^2-(\tilde{J'}_{\rm BC}^{v})^2-(\tilde{J'}_{\rm CA}^{v})^2$. 
Thus, the additional antisymmetric modulation in the up-up-down state is given by $q^5 \cos \phi$, indicating that the modulation of $\cos \phi$ affects the large $\bm{q}$ region in the Brillouin zone. 
Also in these cases in Eqs.~(\ref{eq:F5_kagome_uud_D}) and (\ref{eq:F5_kagome_uud_Ja}), the odd order of the effective DM interaction and the even order of the effective symmetric anisotropic interaction can be a source of the antisymmetric dispersions. 

Such $q^n$ dependence in $\cos \phi$ depends on the model parameters. 
For example, we consider the situation where the breathing parameter for the DM interaction $\gamma_{\rm DM}$ is different from $\gamma$, $\gamma_{\rm DM} \neq \gamma$~\cite{matsumoto2021nonreciprocal}. 
In this case, the $\cos \phi$ dependence appears in $F^{(3)}_{\bm{q}}$ as 
\begin{align}
F^{(3)}_{\bm{q}}=& D  g_1  (\cos q_x -\cos \sqrt{3} q_y) \sin q_x \nonumber \\
&+ D   g_2   \cos \sqrt{3} q_y \sin q_x,    
\end{align}
where $g_1=-24 \gamma_{\rm DM}  (1-\gamma_{\rm DM})D^2+ (\gamma^2-2\gamma+2 \gamma \gamma_{\rm DM}-\gamma_{\rm DM} )J^{\perp 2}$
and $g_2=-48(1+\gamma)(\gamma-\gamma_{\rm DM} ) J^{\perp} J_z$. 
The expression in the form of the effective interaction is omitted due to its length.
Owing to nonzero $g_2$, i.e., $\gamma_{\rm DM} \neq \gamma$, $F^{(3)}_{\bm{q}}$ has the contribution of $q \cos \phi$ in the limit of $|\bm{q}| \to 0$, which means the linear band modulation is found in the small $\bm{q}$ region~\cite{matsumoto2021nonreciprocal}. 

\begin{figure}[t!]
\begin{center}
\includegraphics[width=1.0 \hsize]{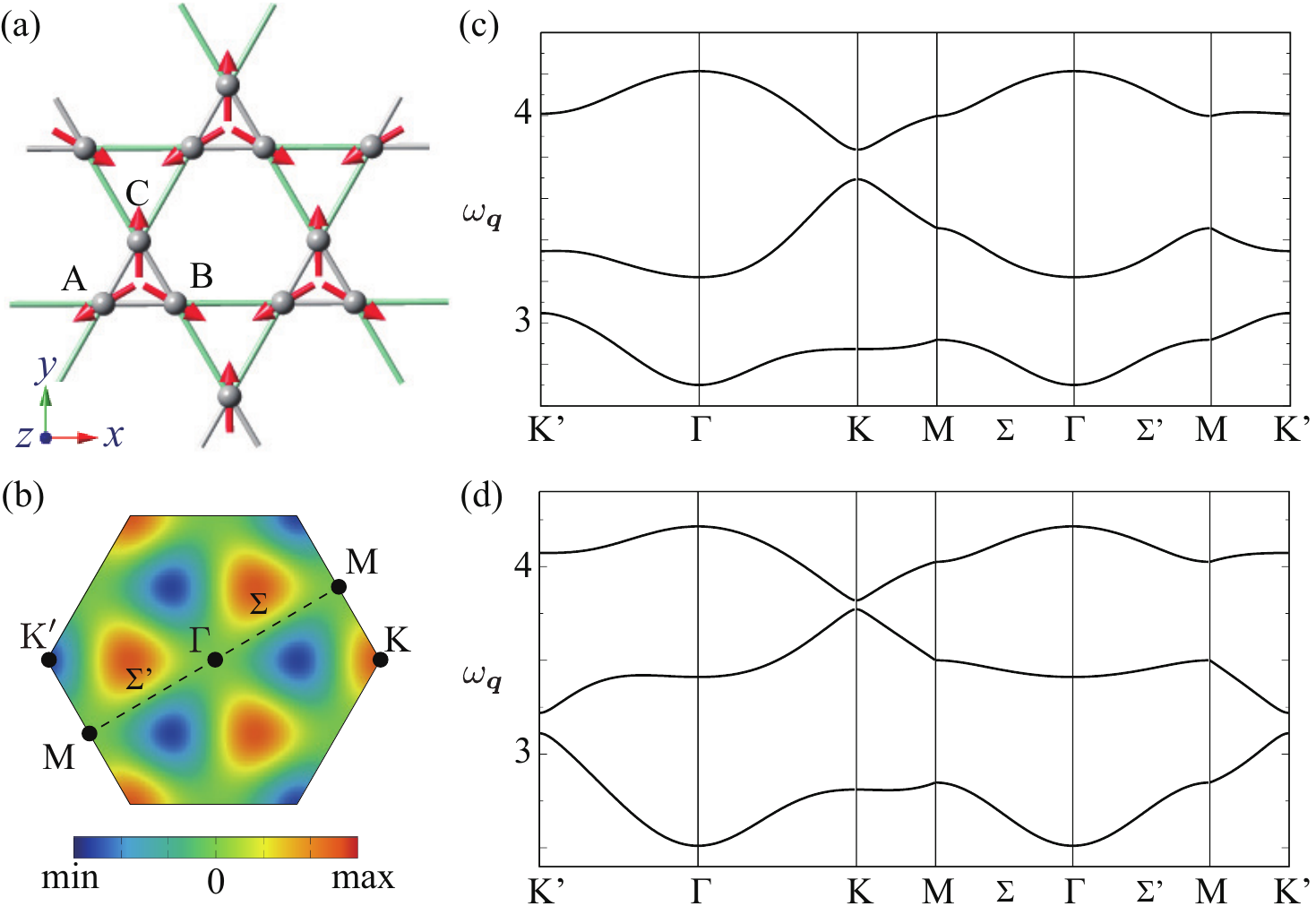} 
\caption{
\label{Fig:BKL_120}
(a) Breathing kagome lattice structure in the absence of the horizontal mirror plane under the polar point group $C_{3v}$. 
The arrows represent the magnetic moments to form the 120$^{\circ}$ antiferromagnetic ordering. 
(b) The first Brillouin zone in (a). 
The color plot represents angle dependence of nonreciprocal magnons characterized by $q_x (q_x^2-3q_y^2)$, which is the same as that in Fig.~\ref{Fig:BKL_ferro}(b). 
(c, d) The magnon band structures under the 120$^{\circ}$ antiferromagnetic ordering for $D'=0.2$ and $J'^a=0$ (c) and $D'=0$ and $J'^a=0.2$ (d).
The other parameters are set as $J^{\perp}=1$, $J^z=0.8$, $D=-0.2$, $J^a=0.5$, and $\gamma=0.5$. 
}
\end{center}
\end{figure}

\subsection{Breathing kagome noncollinear 120$^{\circ}$ antiferromagnets}
\label{sec:Breathing kagome noncollinear 120 antiferromagnets}

\subsubsection{Model}

Finally, we discuss the nonreciprocal magnons in the noncollinear antiferromagnetic state. 
We consider the $120^{\circ}$ antiferromagnetic ordering in the breathing kagome lattice structure in Fig.~\ref{Fig:BKL_120}(a). 
Here, we consider the situation where the horizontal mirror symmetry in the kagome plane is broken owing to the presence of polar field along the $z$ direction, which means that the point group symmetry is lowered to $C_{3v}$. 
Then, the spin Hamiltonian is given by 
\begin{widetext}
\begin{align}
\label{eq:Jmat_BKL_120}
\mathcal{J}^{\triangle}_{\eta\eta'}&=
\begin{pmatrix}
J^{\perp}+J^a \cos \chi_{\eta\eta'}& D-J^a \sin \chi_{\eta\eta'}&-D'\cos \chi_{\eta\eta'}-J'^a \sin \chi_{\eta\eta'} \\
-D-J^a \sin \chi_{\eta\eta'}& J^{\perp}-J^a \cos \chi_{\eta\eta'}&-D'\sin \chi_{\eta\eta'}+J'^a \cos \chi_{\eta\eta'} \\
D'\cos \chi_{\eta\eta'}-J'^a \sin \chi_{\eta\eta'}  &D'\sin \chi_{\eta\eta'}+J'^a \cos \chi_{\eta\eta'} &J^z
\end{pmatrix}, \\
\label{eq:Jmat_BKL2}
\mathcal{J}^{\bigtriangledown}_{\eta\eta'}&=\gamma\mathcal{J}^{\triangle}_{\eta\eta'}, 
\end{align}
\end{widetext}
where $D'$ and $J'^a$ are additional exchange interactions that arise from the horizontal mirror symmetry breaking under the polar field.

The effective interactions in the rotated spin frame are given by
\begin{align}
\tilde{J}_{\rm \eta\eta'}^{\perp}=&-\frac{1}{4} \left(J^{\perp}-2 J^a+2 J^z-\sqrt{3} D\right),\\
\tilde{J}_{\rm \eta\eta'}^{v}=&\frac{1}{4} \left(J^{\perp}-2 J^a+2 J^z-\sqrt{3} D\right),\\
\tilde{J}_{\rm \eta\eta'}^{xy}=&0,\\
\tilde{J}_{\rm \eta\eta'}^{z}=&-\frac{1}{2} \left(J^{\perp}+2 J^a-\sqrt{3} D \right),\\
\tilde{D}_{\rm \eta\eta'}^{z}=&-\frac{1}{2} \left(\sqrt{3} J'^a+D'\right),
\end{align}
where $\eta,\eta'=$ A, B, and C, and we neglect $\tilde{J}_{\rm \eta\eta'}^{zx}$ and $\tilde{D}_{\rm \eta\eta'}^{x}$ owing to the linear spin wave approximation. 
The expressions are the same for the different bonds (A-B, B-C, and C-A) owing to the symmetry.

The $3\times3$ matrices $\mathcal{X}_{\bm{q}}$ and $\mathcal{Y}_{\bm{q}}$ in the Bogoliubov Hamiltonian in momentum space are the same as those in Eqs.~(\ref{eq:Xq_BKL}) and (\ref{eq:Yq_BKL}), respectively. 
Meanwhile, $F_{\eta\eta'{\bm{q}}}$, $G_{\eta\eta'{\bm{q}}}$, and $Z$ have different forms as 
\begin{align}
\label{eq:F_BKL_120}
F_{\eta\eta'{\bm{q}}}
=&
\left[-\frac{J^{\perp}-2 J^a+2 J^z-\sqrt{3} D}{4} +\frac{i(\sqrt{3} J'^a+D')}{2} \right] \nonumber \\
\times 
&\left(e^{i \bm{q} \cdot \bm{\rho}_{\eta\eta'}}
+\gamma
e^{-i \bm{q} \cdot \bm{\rho}_{\eta\eta'}}\right),\\
\label{eq:G_BKL_120}
G_{\eta\eta'{\bm{q}}}
=&
\frac{J^{\perp}-2 J^a+2 J^z-\sqrt{3} D}{4} 
\left(e^{i \bm{q} \cdot \bm{\rho}_{\eta\eta'}}
+\gamma  
e^{-i \bm{q} \cdot \bm{\rho}_{\eta\eta'}}\right),\\
Z=&(1+ \gamma )\left(J^{\perp}+2 J^a-\sqrt{3} D \right). 
\end{align}

\subsubsection{Result}

The 120$^{\circ}$ spin configuration is obtained as a metastable state by taking the exchange model parameters as $J^{\perp}=1$, $J^z=0.8$, $D=-0.2$, $J^a=0.5$, and $\gamma=0.5$. 
Figures~\ref{Fig:BKL_120}(c) and \ref{Fig:BKL_120}(d) show the magnon dispersions under the 120$^{\circ}$ antiferromagnetic ordering along high symmetry lines in the Brillouin zone in Fig.~\ref{Fig:BKL_120}(b). 
The data in Fig.~\ref{Fig:BKL_120}(c) is obtained at $D'=0.2$ and $J'^a=0$ and that in Fig.~\ref{Fig:BKL_120}(d) is at $D'=0$ and $J'^a=0.2$.
Although the interaction tensor under the 120$^{\circ}$ antiferromagnetic ordering is different from that in the ferromagnetic ordering in Eq.~(\ref{eq:Jmat_BKL}), the functional form of the antisymmetric dispersions is the same with each other, which is characterized by $q_x (q_x^2-3q_y^2)$ satisfying threefold rotational symmetry in both cases in Figs.~\ref{Fig:BKL_120}(c) and \ref{Fig:BKL_120}(d).

The lowest-order contribution of $F^{(s)}_{\bm{q}}$ is of third order. 
In the case at $D' \neq 0$ and $J'^a= 0$, $F^{(3)}_{\bm{q}}$ is given by
\begin{align}
F^{(3)}_{\bm{q}}=&-3\gamma (1-\gamma)  D'  f^{3\phi}_{\bm{q}} \nonumber \\
&\times \left[6 J^z \left(J^{\perp}-\sqrt{3} D-2 J^a\right)+D'^2\right],   
\end{align}
and in the case at $D' = 0$ and $J'^a \neq 0$, $F^{(3)}_{\bm{q}}$ is given by
\begin{align}
F^{(3)}_{\bm{q}}=&-9  \gamma (1-\gamma) J'^a f^{3\phi}_{\bm{q}}\nonumber \\ 
&\times \left\{\sqrt{3} \left[2 J^z (J^{\perp}-2 J^a)+(J'^a)^2\right]-6 D J^z\right\}. 
\end{align}
where we omit the expressions for the effective exchange interactions. 
The above results indicates that we obtain the different conditions in terms of the essential model parameters from the ferromagnetic state in Eqs.~(\ref{eq:F_BKL_FM_D}) and (\ref{eq:F_BKL_FM_Ja}): The former are $D'$ and $J'^a$, while the latter are $D$ and $J^a$. 
In this way, our scheme can be applied to noncollinear antiferromagnetic orderings straightforwardly.

\section{Summary}
\label{sec:Summary}

To summarize, we have investigated the microscopic conditions for emergent nonreciprocal magnons on the basis of the model calculations. 
We presented the useful expression in Eqs.~(\ref{eq:Es}) and (\ref{eq:Fs}) to provide essential model parameters for nonreciprocal magnon excitations in an analytical way. 
The method does not require the diagonalization of the bosonic Hamiltonian. 
After presenting the generic results in the one- to four-sublattice cases, we tested the method to four magnetic systems: the ferromagnetic state on the breathing kagome lattice system, the staggered collinear antiferromagnetic state on the honeycomb lattice system, the up-up-down ferrimagnetic state on the breathing kagome lattice system, and the noncollinear 120$^{\circ}$ antiferromagnetic state on the breathing kagome lattice system. 
We found that our scheme extracts the key model parameters, which are well consistent with the result by the direct diagonalization.

The present expression can be applied to any magnetic structures including noncollinear one in the magnetic systems with any symmetric and antisymmetric bilinear exchange interactions. 
In particular, this method has an advantage of obtaining the analytical expressions for the essential model parameters in multisublattice systems with long-period magnetic structures that are difficult to obtain the analytical expressions of the magnon band dispersions. 
Moreover, the systematic analysis provides an insight to construct an effective spin model so as to include essential model parameters in real materials, where targeting materials are easily found by using magnetic structure database, MAGNDATA~\cite{gallego2016magndata}, and cluster multipole analyses~\cite{Suzuki_PhysRevB.99.174407,Yatsushiro_PhysRevB.104.054412}, from the symmetry viewpoint. 
In this way, our result will not only give a deep understanding of nonreciprocal magnon excitations in noncentrosymmetric magnets, such as $\alpha$-Cu$_2$V$_2$O$_7$~\cite{Gitgeatpong_PhysRevB.92.024423,Gitgeatpong_PhysRevLett.119.047201,Gitgeatpong_PhysRevB.95.245119,piyawongwatthana2021formation}, but also be a good indicator to examine the microscopic origin under complicated magnetic orderings. 

\appendix
\section{Expressions of $F^{(s)}_{\bm{q}}$ in three- and four-sublattice cases}
\label{sec:appendix}

In this Appendix, we show the lengthy expressions of $H_{\mu\bm{q}}$ ($\mu=1$-7) in the three-sublattice case in Sec.~\ref{sec:Three-sublattice case} and those of $H'_{\mu\bm{q}}$ ($\mu=1$-7) in the four-sublattice case in Sec.~\ref{sec:Four-sublattice case}. 
\begin{widetext}
For the three-sublattice case, $H_{\mu\bm{q}}$ ($\mu=1$-7) are given by 
\begin{align}
H_{1\bm{q}}=&12\{z_{\rm A} (h^{D{\rm (s)}}_{{\rm AB}\bm{q}} h^{\perp{\rm (as)}}_{{\rm AB}\bm{q}}+h^{D{\rm (s)}}_{{\rm AC}\bm{q}} h^{\perp{\rm (as)}}_{{\rm AC}\bm{q}}-h^{\perp{\rm (s)}}_{{\rm AB}\bm{q}} h^{D{\rm (as)}}_{{\rm AB}\bm{q}}-h^{\perp{\rm (s)}}_{{\rm AC}\bm{q}} h^{D{\rm (as)}}_{{\rm AC}\bm{q}})\nonumber \\
&+z_{\rm B} (h^{D{\rm (s)}}_{{\rm AB}\bm{q}} h^{\perp{\rm (as)}}_{{\rm AB}\bm{q}}+h^{D{\rm (s)}}_{{\rm BC}\bm{q}} h^{\perp{\rm (as)}}_{{\rm BC}\bm{q}}-h^{\perp{\rm (s)}}_{{\rm AB}\bm{q}} h^{D{\rm (as)}}_{{\rm AB}\bm{q}}-h^{\perp{\rm (s)}}_{{\rm BC}\bm{q}} h^{D{\rm (as)}}_{{\rm BC}\bm{q}})\nonumber \\
&+z_{\rm C} (h^{D{\rm (s)}}_{{\rm AC}\bm{q}} h^{\perp{\rm (as)}}_{{\rm AC}\bm{q}}+h^{D{\rm (s)}}_{{\rm BC}\bm{q}} h^{\perp{\rm (as)}}_{{\rm BC}\bm{q}}-h^{\perp{\rm (s)}}_{{\rm AC}\bm{q}} h^{D{\rm (as)}}_{{\rm AC}\bm{q}}-h^{\perp{\rm (s)}}_{{\rm BC}\bm{q}} h^{D{\rm (as)}}_{{\rm BC}\bm{q}})\}
\\
H_{2\bm{q}}=&12(
-h^{D{\rm (s)}}_{{\rm AB}\bm{q}} h^{\perp{\rm (s)}}_{{\rm AC}\bm{q}} h^{\perp{\rm (as)}}_{{\rm BC}\bm{q}}
+h^{D{\rm (s)}}_{{\rm AB}\bm{q}} h^{\perp{\rm (s)}}_{{\rm BC}\bm{q}} h^{\perp{\rm (as)}}_{{\rm AC}\bm{q}}
+h^{D{\rm (s)}}_{{\rm AC}\bm{q}} h^{\perp{\rm (s)}}_{{\rm AB}\bm{q}} h^{\perp{\rm (as)}}_{{\rm BC}\bm{q}}
+h^{D{\rm (s)}}_{{\rm AC}\bm{q}} h^{\perp{\rm (s)}}_{{\rm BC}\bm{q}} h^{\perp{\rm (as)}}_{{\rm AB}\bm{q}}
+h^{D{\rm (s)}}_{{\rm BC}\bm{q}} h^{\perp{\rm (s)}}_{{\rm AB}\bm{q}} h^{\perp{\rm (as)}}_{{\rm AC}\bm{q}} \nonumber \\
&-h^{D{\rm (s)}}_{{\rm BC}\bm{q}} h^{\perp{\rm (s)}}_{{\rm AC}\bm{q}} h^{\perp{\rm (as)}}_{{\rm AB}\bm{q}}
-h^{\perp{\rm (s)}}_{{\rm AB}\bm{q}} h^{\perp{\rm (s)}}_{{\rm AC}\bm{q}} h^{D{\rm (as)}}_{{\rm BC}\bm{q}}
-h^{\perp{\rm (s)}}_{{\rm AB}\bm{q}} h^{\perp{\rm (s)}}_{{\rm BC}\bm{q}} h^{D{\rm (as)}}_{{\rm AC}\bm{q}}
-h^{\perp{\rm (s)}}_{{\rm AC}\bm{q}} h^{\perp{\rm (s)}}_{{\rm BC}\bm{q}} h^{D{\rm (as)}}_{{\rm AB}\bm{q}}
-h^{D{\rm (as)}}_{{\rm AB}\bm{q}} h^{\perp{\rm (as)}}_{{\rm AC}\bm{q}} h^{\perp{\rm (as)}}_{{\rm BC}\bm{q}} \nonumber \\
&+h^{D{\rm (as)}}_{{\rm AC}\bm{q}} h^{\perp{\rm (as)}}_{{\rm AB}\bm{q}} h^{\perp{\rm (as)}}_{{\rm BC}\bm{q}}-h^{D{\rm (as)}}_{{\rm BC}\bm{q}} h^{\perp{\rm (as)}}_{{\rm AB}\bm{q}} h^{\perp{\rm (as)}}_{{\rm AC}\bm{q}})
\\
H_{5\bm{q}}=&12(-h^{D{\rm (s)}}_{{\rm AB}\bm{q}} h^{D{\rm (s)}}_{{\rm AC}\bm{q}} h^{D{\rm (as)}}_{{\rm BC}\bm{q}}+h^{D{\rm (s)}}_{{\rm AB}\bm{q}} h^{D{\rm (s)}}_{{\rm BC}\bm{q}} h^{D{\rm (as)}}_{{\rm AC}\bm{q}}-h^{D{\rm (s)}}_{{\rm AC}\bm{q}} h^{D{\rm (s)}}_{{\rm BC}\bm{q}} h^{D{\rm (as)}}_{{\rm AB}\bm{q}}-h^{D{\rm (as)}}_{{\rm AB}\bm{q}} h^{D{\rm (as)}}_{{\rm AC}\bm{q}} h^{D{\rm (as)}}_{{\rm BC}\bm{q}})
\\
H_{6\bm{q}}=&12\{z_{\rm A} (h^{v{\rm (s)}}_{{\rm AB}\bm{q}} h^{xy{\rm (as)}}_{{\rm AB}\bm{q}}+h^{v{\rm (s)}}_{{\rm AC}\bm{q}} h^{xy{\rm (as)}}_{{\rm AC}\bm{q}}-h^{xy{\rm (s)}}_{{\rm AB}\bm{q}} h^{v{\rm (as)}}_{{\rm AB}\bm{q}}-h^{xy{\rm (s)}}_{{\rm AC}\bm{q}} h^{v{\rm (as)}}_{{\rm AC}\bm{q}})\nonumber \\
&+z_{\rm B} (-h^{v{\rm (s)}}_{{\rm AB}\bm{q}} h^{xy{\rm (as)}}_{{\rm AB}\bm{q}}+h^{v{\rm (s)}}_{{\rm BC}\bm{q}} h^{xy{\rm (as)}}_{{\rm BC}\bm{q}}+h^{xy{\rm (s)}}_{{\rm AB}\bm{q}} h^{v{\rm (as)}}_{{\rm AB}\bm{q}}-h^{xy{\rm (s)}}_{{\rm BC}\bm{q}} h^{v{\rm (as)}}_{{\rm BC}\bm{q}})\nonumber \\
&+z_{\rm C} (-h^{v{\rm (s)}}_{{\rm AC}\bm{q}} h^{xy{\rm (as)}}_{{\rm AC}\bm{q}}-h^{v{\rm (s)}}_{{\rm BC}\bm{q}} h^{xy{\rm (as)}}_{{\rm BC}\bm{q}}+h^{xy{\rm (s)}}_{{\rm AC}\bm{q}} h^{v{\rm (as)}}_{{\rm AC}\bm{q}}+h^{xy{\rm (s)}}_{{\rm BC}\bm{q}} h^{v{\rm (as)}}_{{\rm BC}\bm{q}})\}
\\
H_{7\bm{q}}=&12(
h^{\perp{\rm (s)}}_{{\rm AB}\bm{q}} h^{v{\rm (s)}}_{{\rm BC}\bm{q}} h^{xy{\rm (as)}}_{{\rm AC}\bm{q}}
+h^{\perp{\rm (s)}}_{{\rm AC}\bm{q}} h^{v{\rm (s)}}_{{\rm BC}\bm{q}} h^{xy{\rm (as)}}_{{\rm AB}\bm{q}}
-h^{v{\rm (s)}}_{{\rm BC}\bm{q}} h^{xy{\rm (s)}}_{{\rm AB}\bm{q}} h^{\perp{\rm (as)}}_{{\rm AC}\bm{q}}
-h^{v{\rm (s)}}_{{\rm BC}\bm{q}} h^{xy{\rm (s)}}_{{\rm AC}\bm{q}} h^{\perp{\rm (as)}}_{{\rm AB}\bm{q}}
+h^{\perp{\rm (s)}}_{{\rm AB}\bm{q}} h^{v{\rm (s)}}_{{\rm AC}\bm{q}} h^{xy{\rm (as)}}_{{\rm BC}\bm{q}} \nonumber \\
&-h^{\perp{\rm (s)}}_{{\rm BC}\bm{q}} h^{v{\rm (s)}}_{{\rm AC}\bm{q}} h^{xy{\rm (as)}}_{{\rm AB}\bm{q}}
-h^{v{\rm (s)}}_{{\rm AC}\bm{q}} h^{xy{\rm (s)}}_{{\rm AB}\bm{q}} h^{\perp{\rm (as)}}_{{\rm BC}\bm{q}}
+h^{v{\rm (s)}}_{{\rm AC}\bm{q}} h^{xy{\rm (s)}}_{{\rm BC}\bm{q}} h^{\perp{\rm (as)}}_{{\rm AB}\bm{q}}
-h^{\perp{\rm (s)}}_{{\rm AB}\bm{q}} h^{xy{\rm (s)}}_{{\rm AC}\bm{q}} h^{v{\rm (as)}}_{{\rm BC}\bm{q}}
-h^{\perp{\rm (s)}}_{{\rm AB}\bm{q}} h^{xy{\rm (s)}}_{{\rm BC}\bm{q}} h^{v{\rm (as)}}_{{\rm AC}\bm{q}} \nonumber \\
&-h^{\perp{\rm (s)}}_{{\rm AC}\bm{q}} h^{v{\rm (s)}}_{{\rm AB}\bm{q}} h^{xy{\rm (as)}}_{{\rm BC}\bm{q}}
+h^{\perp{\rm (s)}}_{{\rm AC}\bm{q}} h^{xy{\rm (s)}}_{{\rm AB}\bm{q}} h^{v{\rm (as)}}_{{\rm BC}\bm{q}}
-h^{\perp{\rm (s)}}_{{\rm AC}\bm{q}} h^{xy{\rm (s)}}_{{\rm BC}\bm{q}} h^{v{\rm (as)}}_{{\rm AB}\bm{q}}
-h^{\perp{\rm (s)}}_{{\rm BC}\bm{q}} h^{v{\rm (s)}}_{{\rm AB}\bm{q}} h^{xy{\rm (as)}}_{{\rm AC}\bm{q}}
+h^{\perp{\rm (s)}}_{{\rm BC}\bm{q}} h^{xy{\rm (s)}}_{{\rm AB}\bm{q}} h^{v{\rm (as)}}_{{\rm AC}\bm{q}} \nonumber \\
&+h^{\perp{\rm (s)}}_{{\rm BC}\bm{q}} h^{xy{\rm (s)}}_{{\rm AC}\bm{q}} h^{v{\rm (as)}}_{{\rm AB}\bm{q}}
+h^{v{\rm (s)}}_{{\rm AB}\bm{q}} h^{xy{\rm (s)}}_{{\rm AC}\bm{q}} h^{\perp{\rm (as)}}_{{\rm BC}\bm{q}}
+h^{v{\rm (s)}}_{{\rm AB}\bm{q}} h^{xy{\rm (s)}}_{{\rm BC}\bm{q}} h^{\perp{\rm (as)}}_{{\rm AC}\bm{q}}
+h^{\perp{\rm (as)}}_{{\rm AB}\bm{q}} h^{v{\rm (as)}}_{{\rm AC}\bm{q}} h^{xy{\rm (as)}}_{{\rm BC}\bm{q}}
-h^{\perp{\rm (as)}}_{{\rm AB}\bm{q}} h^{v{\rm (as)}}_{{\rm BC}\bm{q}} h^{xy{\rm (as)}}_{{\rm AC}\bm{q}}\nonumber \\
&-h^{\perp{\rm (as)}}_{{\rm AC}\bm{q}} h^{v{\rm (as)}}_{{\rm AB}\bm{q}} h^{xy{\rm (as)}}_{{\rm BC}\bm{q}} 
+h^{\perp{\rm (as)}}_{{\rm AC}\bm{q}} h^{v{\rm (as)}}_{{\rm BC}\bm{q}} h^{xy{\rm (as)}}_{{\rm AB}\bm{q}}
+h^{\perp{\rm (as)}}_{{\rm BC}\bm{q}} h^{v{\rm (as)}}_{{\rm AB}\bm{q}} h^{xy{\rm (as)}}_{{\rm AC}\bm{q}}
-h^{\perp{\rm (as)}}_{{\rm BC}\bm{q}} h^{v{\rm (as)}}_{{\rm AC}\bm{q}} h^{xy{\rm (as)}}_{{\rm AB}\bm{q}}),
\end{align}
where $H_{3\bm{q}}$ and $H_{4\bm{q}}$ are obtained by replacing the superscript $\perp$ in $H_{2\bm{q}}$ with $v$ and $xy$, respectively, and multiplying $-1$.

For the four-sublattice case, $H'_{\mu\bm{q}}$ ($\mu=1$-7) are given by 
\begin{align}
H'_{1\bm{q}}=&12\{
z_{\rm A} (h^{D{\rm (s)}}_{{\rm AB}\bm{q}} h^{\perp{\rm (as)}}_{{\rm AB}\bm{q}}+h^{D{\rm (s)}}_{{\rm AC}\bm{q}} h^{\perp{\rm (as)}}_{{\rm AC}\bm{q}}+h^{D{\rm (s)}}_{{\rm AD}\bm{q}} h^{\perp{\rm (as)}}_{{\rm AD}\bm{q}}-h^{\perp{\rm (s)}}_{{\rm AB}\bm{q}} h^{D{\rm (as)}}_{{\rm AB}\bm{q}}-h^{\perp{\rm (s)}}_{{\rm AC}\bm{q}} h^{D{\rm (as)}}_{{\rm AC}\bm{q}}-h^{\perp{\rm (s)}}_{{\rm AD}\bm{q}} h^{D{\rm (as)}}_{{\rm AD}\bm{q}})\nonumber \\
&+z_{\rm B} (h^{D{\rm (s)}}_{{\rm AB}\bm{q}} h^{\perp{\rm (as)}}_{{\rm AB}\bm{q}}+h^{D{\rm (s)}}_{{\rm BC}\bm{q}} h^{\perp{\rm (as)}}_{{\rm BC}\bm{q}}+h^{D{\rm (s)}}_{{\rm BD}\bm{q}} h^{\perp{\rm (as)}}_{{\rm BD}\bm{q}}-h^{\perp{\rm (s)}}_{{\rm AB}\bm{q}} h^{D{\rm (as)}}_{{\rm AB}\bm{q}}-h^{\perp{\rm (s)}}_{{\rm BC}\bm{q}} h^{D{\rm (as)}}_{{\rm BC}\bm{q}}-h^{\perp{\rm (s)}}_{{\rm BD}\bm{q}} h^{D{\rm (as)}}_{{\rm BD}\bm{q}})\nonumber \\
&+z_{\rm C} (h^{D{\rm (s)}}_{{\rm AC}\bm{q}} h^{\perp{\rm (as)}}_{{\rm AC}\bm{q}}+h^{D{\rm (s)}}_{{\rm BC}\bm{q}} h^{\perp{\rm (as)}}_{{\rm BC}\bm{q}}+h^{D{\rm (s)}}_{{\rm CD}\bm{q}} h^{\perp{\rm (as)}}_{{\rm CD}\bm{q}}-h^{\perp{\rm (s)}}_{{\rm AC}\bm{q}} h^{D{\rm (as)}}_{{\rm AC}\bm{q}}-h^{\perp{\rm (s)}}_{{\rm BC}\bm{q}} h^{D{\rm (as)}}_{{\rm BC}\bm{q}}-h^{\perp{\rm (s)}}_{{\rm CD}\bm{q}} h^{D{\rm (as)}}_{{\rm CD}\bm{q}})\nonumber \\
&+z_{\rm D} (h^{D{\rm (s)}}_{{\rm AD}\bm{q}} h^{\perp{\rm (as)}}_{{\rm AD}\bm{q}}+h^{D{\rm (s)}}_{{\rm BD}\bm{q}} h^{\perp{\rm (as)}}_{{\rm BD}\bm{q}}+h^{D{\rm (s)}}_{{\rm CD}\bm{q}} h^{\perp{\rm (as)}}_{{\rm CD}\bm{q}}-h^{\perp{\rm (s)}}_{{\rm AD}\bm{q}} h^{D{\rm (as)}}_{{\rm AD}\bm{q}}-h^{\perp{\rm (s)}}_{{\rm BD}\bm{q}} h^{D{\rm (as)}}_{{\rm BD}\bm{q}}-h^{\perp{\rm (s)}}_{{\rm CD}\bm{q}} h^{D{\rm (as)}}_{{\rm CD}\bm{q}})\}
\\
H'_{2\bm{q}}=&12(
-h^{D{\rm (s)}}_{{\rm AB}\bm{q}} h^{\perp{\rm (s)}}_{{\rm AC}\bm{q}} h^{\perp{\rm (as)}}_{{\rm BC}\bm{q}}
-h^{D{\rm (s)}}_{{\rm AB}\bm{q}} h^{\perp{\rm (s)}}_{{\rm AD}\bm{q}} h^{\perp{\rm (as)}}_{{\rm BD}\bm{q}}
+h^{D{\rm (s)}}_{{\rm AB}\bm{q}} h^{\perp{\rm (s)}}_{{\rm BC}\bm{q}} h^{\perp{\rm (as)}}_{{\rm AC}\bm{q}}
+h^{D{\rm (s)}}_{{\rm AB}\bm{q}} h^{\perp{\rm (s)}}_{{\rm BD}\bm{q}} h^{\perp{\rm (as)}}_{{\rm AD}\bm{q}}
+h^{D{\rm (s)}}_{{\rm AC}\bm{q}} h^{\perp{\rm (s)}}_{{\rm AB}\bm{q}} h^{\perp{\rm (as)}}_{{\rm BC}\bm{q}} \nonumber \\
&-h^{D{\rm (s)}}_{{\rm AC}\bm{q}} h^{\perp{\rm (s)}}_{{\rm AD}\bm{q}} h^{\perp{\rm (as)}}_{{\rm CD}\bm{q}}
+h^{D{\rm (s)}}_{{\rm AC}\bm{q}} h^{\perp{\rm (s)}}_{{\rm BC}\bm{q}} h^{\perp{\rm (as)}}_{{\rm AB}\bm{q}}
+h^{D{\rm (s)}}_{{\rm AC}\bm{q}} h^{\perp{\rm (s)}}_{{\rm CD}\bm{q}} h^{\perp{\rm (as)}}_{{\rm AD}\bm{q}}
+h^{D{\rm (s)}}_{{\rm AD}\bm{q}} h^{\perp{\rm (s)}}_{{\rm AB}\bm{q}} h^{\perp{\rm (as)}}_{{\rm BD}\bm{q}}
+h^{D{\rm (s)}}_{{\rm AD}\bm{q}} h^{\perp{\rm (s)}}_{{\rm AC}\bm{q}} h^{\perp{\rm (as)}}_{{\rm CD}\bm{q}} \nonumber \\
&+h^{D{\rm (s)}}_{{\rm AD}\bm{q}} h^{\perp{\rm (s)}}_{{\rm BD}\bm{q}} h^{\perp{\rm (as)}}_{{\rm AB}\bm{q}}
+h^{D{\rm (s)}}_{{\rm AD}\bm{q}} h^{\perp{\rm (s)}}_{{\rm CD}\bm{q}} h^{\perp{\rm (as)}}_{{\rm AC}\bm{q}}
+h^{D{\rm (s)}}_{{\rm BC}\bm{q}} h^{\perp{\rm (s)}}_{{\rm AB}\bm{q}} h^{\perp{\rm (as)}}_{{\rm AC}\bm{q}}
-h^{D{\rm (s)}}_{{\rm BC}\bm{q}} h^{\perp{\rm (s)}}_{{\rm AC}\bm{q}} h^{\perp{\rm (as)}}_{{\rm AB}\bm{q}}
-h^{D{\rm (s)}}_{{\rm BC}\bm{q}} h^{\perp{\rm (s)}}_{{\rm BD}\bm{q}} h^{\perp{\rm (as)}}_{{\rm CD}\bm{q}} \nonumber \\
&+h^{D{\rm (s)}}_{{\rm BC}\bm{q}} h^{\perp{\rm (s)}}_{{\rm CD}\bm{q}} h^{\perp{\rm (as)}}_{{\rm BD}\bm{q}}
+h^{D{\rm (s)}}_{{\rm BD}\bm{q}} h^{\perp{\rm (s)}}_{{\rm AB}\bm{q}} h^{\perp{\rm (as)}}_{{\rm AD}\bm{q}}
-h^{D{\rm (s)}}_{{\rm BD}\bm{q}} h^{\perp{\rm (s)}}_{{\rm AD}\bm{q}} h^{\perp{\rm (as)}}_{{\rm AB}\bm{q}}
+h^{D{\rm (s)}}_{{\rm BD}\bm{q}} h^{\perp{\rm (s)}}_{{\rm BC}\bm{q}} h^{\perp{\rm (as)}}_{{\rm CD}\bm{q}}
+h^{D{\rm (s)}}_{{\rm BD}\bm{q}} h^{\perp{\rm (s)}}_{{\rm CD}\bm{q}} h^{\perp{\rm (as)}}_{{\rm BC}\bm{q}}\nonumber \\
&+h^{D{\rm (s)}}_{{\rm CD}\bm{q}} h^{\perp{\rm (s)}}_{{\rm AC}\bm{q}} h^{\perp{\rm (as)}}_{{\rm AD}\bm{q}} 
-h^{D{\rm (s)}}_{{\rm CD}\bm{q}} h^{\perp{\rm (s)}}_{{\rm AD}\bm{q}} h^{\perp{\rm (as)}}_{{\rm AC}\bm{q}}
+h^{D{\rm (s)}}_{{\rm CD}\bm{q}} h^{\perp{\rm (s)}}_{{\rm BC}\bm{q}} h^{\perp{\rm (as)}}_{{\rm BD}\bm{q}}
-h^{D{\rm (s)}}_{{\rm CD}\bm{q}} h^{\perp{\rm (s)}}_{{\rm BD}\bm{q}} h^{\perp{\rm (as)}}_{{\rm BC}\bm{q}}
-h^{\perp{\rm (s)}}_{{\rm AB}\bm{q}} h^{\perp{\rm (s)}}_{{\rm AC}\bm{q}} h^{D{\rm (as)}}_{{\rm BC}\bm{q}}\nonumber \\
&-h^{\perp{\rm (s)}}_{{\rm AB}\bm{q}} h^{\perp{\rm (s)}}_{{\rm AD}\bm{q}} h^{D{\rm (as)}}_{{\rm BD}\bm{q}}
-h^{\perp{\rm (s)}}_{{\rm AB}\bm{q}} h^{\perp{\rm (s)}}_{{\rm BC}\bm{q}} h^{D{\rm (as)}}_{{\rm AC}\bm{q}} 
-h^{\perp{\rm (s)}}_{{\rm AB}\bm{q}} h^{\perp{\rm (s)}}_{{\rm BD}\bm{q}} h^{D{\rm (as)}}_{{\rm AD}\bm{q}}
-h^{\perp{\rm (s)}}_{{\rm AC}\bm{q}} h^{\perp{\rm (s)}}_{{\rm AD}\bm{q}} h^{D{\rm (as)}}_{{\rm CD}\bm{q}}
-h^{\perp{\rm (s)}}_{{\rm AC}\bm{q}} h^{\perp{\rm (s)}}_{{\rm BC}\bm{q}} h^{D{\rm (as)}}_{{\rm AB}\bm{q}}\nonumber \\
&-h^{\perp{\rm (s)}}_{{\rm AC}\bm{q}} h^{\perp{\rm (s)}}_{{\rm CD}\bm{q}} h^{D{\rm (as)}}_{{\rm AD}\bm{q}}
-h^{\perp{\rm (s)}}_{{\rm AD}\bm{q}} h^{\perp{\rm (s)}}_{{\rm BD}\bm{q}} h^{D{\rm (as)}}_{{\rm AB}\bm{q}}
-h^{\perp{\rm (s)}}_{{\rm AD}\bm{q}} h^{\perp{\rm (s)}}_{{\rm CD}\bm{q}} h^{D{\rm (as)}}_{{\rm AC}\bm{q}}
-h^{\perp{\rm (s)}}_{{\rm BC}\bm{q}} h^{\perp{\rm (s)}}_{{\rm BD}\bm{q}} h^{D{\rm (as)}}_{{\rm CD}\bm{q}}
-h^{\perp{\rm (s)}}_{{\rm BC}\bm{q}} h^{\perp{\rm (s)}}_{{\rm CD}\bm{q}} h^{D{\rm (as)}}_{{\rm BD}\bm{q}}\nonumber \\
&-h^{\perp{\rm (s)}}_{{\rm BD}\bm{q}} h^{\perp{\rm (s)}}_{{\rm CD}\bm{q}} h^{D{\rm (as)}}_{{\rm BC}\bm{q}}
-h^{D{\rm (as)}}_{{\rm AB}\bm{q}} h^{\perp{\rm (as)}}_{{\rm AC}\bm{q}} h^{\perp{\rm (as)}}_{{\rm BC}\bm{q}}
-h^{D{\rm (as)}}_{{\rm AB}\bm{q}} h^{\perp{\rm (as)}}_{{\rm AD}\bm{q}} h^{\perp{\rm (as)}}_{{\rm BD}\bm{q}}
+h^{D{\rm (as)}}_{{\rm AC}\bm{q}} h^{\perp{\rm (as)}}_{{\rm AB}\bm{q}} h^{\perp{\rm (as)}}_{{\rm BC}\bm{q}}
-h^{D{\rm (as)}}_{{\rm AC}\bm{q}} h^{\perp{\rm (as)}}_{{\rm AD}\bm{q}} h^{\perp{\rm (as)}}_{{\rm CD}\bm{q}} \nonumber \\
&+h^{D{\rm (as)}}_{{\rm AD}\bm{q}} h^{\perp{\rm (as)}}_{{\rm AB}\bm{q}} h^{\perp{\rm (as)}}_{{\rm BD}\bm{q}}
+h^{D{\rm (as)}}_{{\rm AD}\bm{q}} h^{\perp{\rm (as)}}_{{\rm AC}\bm{q}} h^{\perp{\rm (as)}}_{{\rm CD}\bm{q}}
-h^{D{\rm (as)}}_{{\rm BC}\bm{q}} h^{\perp{\rm (as)}}_{{\rm AB}\bm{q}} h^{\perp{\rm (as)}}_{{\rm AC}\bm{q}}
-h^{D{\rm (as)}}_{{\rm BC}\bm{q}} h^{\perp{\rm (as)}}_{{\rm BD}\bm{q}} h^{\perp{\rm (as)}}_{{\rm CD}\bm{q}} 
-h^{D{\rm (as)}}_{{\rm BD}\bm{q}} h^{\perp{\rm (as)}}_{{\rm AB}\bm{q}} h^{\perp{\rm (as)}}_{{\rm AD}\bm{q}}\nonumber \\
&+h^{D{\rm (as)}}_{{\rm BD}\bm{q}} h^{\perp{\rm (as)}}_{{\rm BC}\bm{q}} h^{\perp{\rm (as)}}_{{\rm CD}\bm{q}}-h^{D{\rm (as)}}_{{\rm CD}\bm{q}} h^{\perp{\rm (as)}}_{{\rm AC}\bm{q}} h^{\perp{\rm (as)}}_{{\rm AD}\bm{q}}-h^{D{\rm (as)}}_{{\rm CD}\bm{q}} h^{\perp{\rm (as)}}_{{\rm BC}\bm{q}} h^{\perp{\rm (as)}}_{{\rm BD}\bm{q}})
\\
H'_{5\bm{q}}=&12(
-h^{D{\rm (s)}}_{{\rm AB}\bm{q}} h^{D{\rm (s)}}_{{\rm AC}\bm{q}} h^{D{\rm (as)}}_{{\rm BC}\bm{q}}
-h^{D{\rm (s)}}_{{\rm AB}\bm{q}} h^{D{\rm (s)}}_{{\rm AD}\bm{q}} h^{D{\rm (as)}}_{{\rm BD}\bm{q}}
+h^{D{\rm (s)}}_{{\rm AB}\bm{q}} h^{D{\rm (s)}}_{{\rm BC}\bm{q}} h^{D{\rm (as)}}_{{\rm AC}\bm{q}}
+h^{D{\rm (s)}}_{{\rm AB}\bm{q}} h^{D{\rm (s)}}_{{\rm BD}\bm{q}} h^{D{\rm (as)}}_{{\rm AD}\bm{q}}
-h^{D{\rm (s)}}_{{\rm AC}\bm{q}} h^{D{\rm (s)}}_{{\rm AD}\bm{q}} h^{D{\rm (as)}}_{{\rm CD}\bm{q}} \nonumber \\
&-h^{D{\rm (s)}}_{{\rm AC}\bm{q}} h^{D{\rm (s)}}_{{\rm BC}\bm{q}} h^{D{\rm (as)}}_{{\rm AB}\bm{q}}
+h^{D{\rm (s)}}_{{\rm AC}\bm{q}} h^{D{\rm (s)}}_{{\rm CD}\bm{q}} h^{D{\rm (as)}}_{{\rm AD}\bm{q}}
-h^{D{\rm (s)}}_{{\rm AD}\bm{q}} h^{D{\rm (s)}}_{{\rm BD}\bm{q}} h^{D{\rm (as)}}_{{\rm AB}\bm{q}}
-h^{D{\rm (s)}}_{{\rm AD}\bm{q}} h^{D{\rm (s)}}_{{\rm CD}\bm{q}} h^{D{\rm (as)}}_{{\rm AC}\bm{q}}
-h^{D{\rm (s)}}_{{\rm BC}\bm{q}} h^{D{\rm (s)}}_{{\rm BD}\bm{q}} h^{D{\rm (as)}}_{{\rm CD}\bm{q}} \nonumber \\
&+h^{D{\rm (s)}}_{{\rm BC}\bm{q}} h^{D{\rm (s)}}_{{\rm CD}\bm{q}} h^{D{\rm (as)}}_{{\rm BD}\bm{q}}
-h^{D{\rm (s)}}_{{\rm BD}\bm{q}} h^{D{\rm (s)}}_{{\rm CD}\bm{q}} h^{D{\rm (as)}}_{{\rm BC}\bm{q}}
-h^{D{\rm (as)}}_{{\rm AB}\bm{q}} h^{D{\rm (as)}}_{{\rm AC}\bm{q}} h^{D{\rm (as)}}_{{\rm BC}\bm{q}}
-h^{D{\rm (as)}}_{{\rm AB}\bm{q}} h^{D{\rm (as)}}_{{\rm AD}\bm{q}} h^{D{\rm (as)}}_{{\rm BD}\bm{q}}
-h^{D{\rm (as)}}_{{\rm AC}\bm{q}} h^{D{\rm (as)}}_{{\rm AD}\bm{q}} h^{D{\rm (as)}}_{{\rm CD}\bm{q}} \nonumber \\
& -h^{D{\rm (as)}}_{{\rm BC}\bm{q}} h^{D{\rm (as)}}_{{\rm BD}\bm{q}} h^{D{\rm (as)}}_{{\rm CD}\bm{q}})
\\
H'_{6\bm{q}}=&12
\{z_{\rm A} (h^{v{\rm (s)}}_{{\rm AB}\bm{q}} h^{xy{\rm (as)}}_{{\rm AB}\bm{q}}+h^{v{\rm (s)}}_{{\rm AC}\bm{q}} h^{xy{\rm (as)}}_{{\rm AC}\bm{q}}+h^{v{\rm (s)}}_{{\rm AD}\bm{q}} h^{xy{\rm (as)}}_{{\rm AD}\bm{q}}-h^{xy{\rm (s)}}_{{\rm AB}\bm{q}} h^{v{\rm (as)}}_{{\rm AB}\bm{q}}-h^{xy{\rm (s)}}_{{\rm AC}\bm{q}} h^{v{\rm (as)}}_{{\rm AC}\bm{q}}-h^{xy{\rm (s)}}_{{\rm AD}\bm{q}} h^{v{\rm (as)}}_{{\rm AD}\bm{q}})
\nonumber \\
&+z_{\rm B} (-h^{v{\rm (s)}}_{{\rm AB}\bm{q}} h^{xy{\rm (as)}}_{{\rm AB}\bm{q}}+h^{v{\rm (s)}}_{{\rm BC}\bm{q}} h^{xy{\rm (as)}}_{{\rm BC}\bm{q}}+h^{v{\rm (s)}}_{{\rm BD}\bm{q}} h^{xy{\rm (as)}}_{{\rm BD}\bm{q}}+h^{xy{\rm (s)}}_{{\rm AB}\bm{q}} h^{v{\rm (as)}}_{{\rm AB}\bm{q}}-h^{xy{\rm (s)}}_{{\rm BC}\bm{q}} h^{v{\rm (as)}}_{{\rm BC}\bm{q}}-h^{xy{\rm (s)}}_{{\rm BD}\bm{q}} h^{v{\rm (as)}}_{{\rm BD}\bm{q}}) \nonumber \\ 
&+z_{\rm C} (-h^{v{\rm (s)}}_{{\rm AC}\bm{q}} h^{xy{\rm (as)}}_{{\rm AC}\bm{q}}-h^{v{\rm (s)}}_{{\rm BC}\bm{q}} h^{xy{\rm (as)}}_{{\rm BC}\bm{q}}+h^{v{\rm (s)}}_{{\rm CD}\bm{q}} h^{xy{\rm (as)}}_{{\rm CD}\bm{q}}+h^{xy{\rm (s)}}_{{\rm AC}\bm{q}} h^{v{\rm (as)}}_{{\rm AC}\bm{q}}+h^{xy{\rm (s)}}_{{\rm BC}\bm{q}} h^{v{\rm (as)}}_{{\rm BC}\bm{q}}-h^{xy{\rm (s)}}_{{\rm CD}\bm{q}} h^{v{\rm (as)}}_{{\rm CD}\bm{q}}) \nonumber \\
&+z_{\rm D} (-h^{v{\rm (s)}}_{{\rm AD}\bm{q}} h^{xy{\rm (as)}}_{{\rm AD}\bm{q}}-h^{v{\rm (s)}}_{{\rm BD}\bm{q}} h^{xy{\rm (as)}}_{{\rm BD}\bm{q}}-h^{v{\rm (s)}}_{{\rm CD}\bm{q}} h^{xy{\rm (as)}}_{{\rm CD}\bm{q}}+h^{xy{\rm (s)}}_{{\rm AD}\bm{q}} h^{v{\rm (as)}}_{{\rm AD}\bm{q}}+h^{xy{\rm (s)}}_{{\rm BD}\bm{q}} h^{v{\rm (as)}}_{{\rm BD}\bm{q}}+h^{xy{\rm (s)}}_{{\rm CD}\bm{q}} h^{v{\rm (as)}}_{{\rm CD}\bm{q}})\} \\
H'_{7\bm{q}}=&12(
h^{\perp{\rm (s)}}_{{\rm AB}\bm{q}} h^{v{\rm (s)}}_{{\rm AC}\bm{q}} h^{xy{\rm (as)}}_{{\rm BC}\bm{q}}
+h^{\perp{\rm (s)}}_{{\rm AB}\bm{q}} h^{v{\rm (s)}}_{{\rm AD}\bm{q}} h^{xy{\rm (as)}}_{{\rm BD}\bm{q}}
+h^{\perp{\rm (s)}}_{{\rm AB}\bm{q}} h^{v{\rm (s)}}_{{\rm BC}\bm{q}} h^{xy{\rm (as)}}_{{\rm AC}\bm{q}}
+h^{\perp{\rm (s)}}_{{\rm AB}\bm{q}} h^{v{\rm (s)}}_{{\rm BD}\bm{q}} h^{xy{\rm (as)}}_{{\rm AD}\bm{q}}
-h^{\perp{\rm (s)}}_{{\rm AB}\bm{q}} h^{xy{\rm (s)}}_{{\rm AC}\bm{q}} h^{v{\rm (as)}}_{{\rm BC}\bm{q}} \nonumber \\
&-h^{\perp{\rm (s)}}_{{\rm AB}\bm{q}} h^{xy{\rm (s)}}_{{\rm AD}\bm{q}} h^{v{\rm (as)}}_{{\rm BD}\bm{q}}
-h^{\perp{\rm (s)}}_{{\rm AB}\bm{q}} h^{xy{\rm (s)}}_{{\rm BC}\bm{q}} h^{v{\rm (as)}}_{{\rm AC}\bm{q}}
-h^{\perp{\rm (s)}}_{{\rm AB}\bm{q}} h^{xy{\rm (s)}}_{{\rm BD}\bm{q}} h^{v{\rm (as)}}_{{\rm AD}\bm{q}}
-h^{\perp{\rm (s)}}_{{\rm AC}\bm{q}} h^{v{\rm (s)}}_{{\rm AB}\bm{q}} h^{xy{\rm (as)}}_{{\rm BC}\bm{q}}
+h^{\perp{\rm (s)}}_{{\rm AC}\bm{q}} h^{v{\rm (s)}}_{{\rm AD}\bm{q}} h^{xy{\rm (as)}}_{{\rm CD}\bm{q}} \nonumber \\
&+h^{\perp{\rm (s)}}_{{\rm AC}\bm{q}} h^{v{\rm (s)}}_{{\rm BC}\bm{q}} h^{xy{\rm (as)}}_{{\rm AB}\bm{q}}
+h^{\perp{\rm (s)}}_{{\rm AC}\bm{q}} h^{v{\rm (s)}}_{{\rm CD}\bm{q}} h^{xy{\rm (as)}}_{{\rm AD}\bm{q}}
+h^{\perp{\rm (s)}}_{{\rm AC}\bm{q}} h^{xy{\rm (s)}}_{{\rm AB}\bm{q}} h^{v{\rm (as)}}_{{\rm BC}\bm{q}}
-h^{\perp{\rm (s)}}_{{\rm AC}\bm{q}} h^{xy{\rm (s)}}_{{\rm AD}\bm{q}} h^{v{\rm (as)}}_{{\rm CD}\bm{q}}
-h^{\perp{\rm (s)}}_{{\rm AC}\bm{q}} h^{xy{\rm (s)}}_{{\rm BC}\bm{q}} h^{v{\rm (as)}}_{{\rm AB}\bm{q}} \nonumber \\
&-h^{\perp{\rm (s)}}_{{\rm AC}\bm{q}} h^{xy{\rm (s)}}_{{\rm CD}\bm{q}} h^{v{\rm (as)}}_{{\rm AD}\bm{q}}
-h^{\perp{\rm (s)}}_{{\rm AD}\bm{q}} h^{v{\rm (s)}}_{{\rm AB}\bm{q}} h^{xy{\rm (as)}}_{{\rm BD}\bm{q}}
-h^{\perp{\rm (s)}}_{{\rm AD}\bm{q}} h^{v{\rm (s)}}_{{\rm AC}\bm{q}} h^{xy{\rm (as)}}_{{\rm CD}\bm{q}}
+h^{\perp{\rm (s)}}_{{\rm AD}\bm{q}} h^{v{\rm (s)}}_{{\rm BD}\bm{q}} h^{xy{\rm (as)}}_{{\rm AB}\bm{q}}
+h^{\perp{\rm (s)}}_{{\rm AD}\bm{q}} h^{v{\rm (s)}}_{{\rm CD}\bm{q}} h^{xy{\rm (as)}}_{{\rm AC}\bm{q}} \nonumber \\
&+h^{\perp{\rm (s)}}_{{\rm AD}\bm{q}} h^{xy{\rm (s)}}_{{\rm AB}\bm{q}} h^{v{\rm (as)}}_{{\rm BD}\bm{q}}
+h^{\perp{\rm (s)}}_{{\rm AD}\bm{q}} h^{xy{\rm (s)}}_{{\rm AC}\bm{q}} h^{v{\rm (as)}}_{{\rm CD}\bm{q}}
-h^{\perp{\rm (s)}}_{{\rm AD}\bm{q}} h^{xy{\rm (s)}}_{{\rm BD}\bm{q}} h^{v{\rm (as)}}_{{\rm AB}\bm{q}}
-h^{\perp{\rm (s)}}_{{\rm AD}\bm{q}} h^{xy{\rm (s)}}_{{\rm CD}\bm{q}} h^{v{\rm (as)}}_{{\rm AC}\bm{q}}
-h^{\perp{\rm (s)}}_{{\rm BC}\bm{q}} h^{v{\rm (s)}}_{{\rm AB}\bm{q}} h^{xy{\rm (as)}}_{{\rm AC}\bm{q}} \nonumber \\
&-h^{\perp{\rm (s)}}_{{\rm BC}\bm{q}} h^{v{\rm (s)}}_{{\rm AC}\bm{q}} h^{xy{\rm (as)}}_{{\rm AB}\bm{q}}
+h^{\perp{\rm (s)}}_{{\rm BC}\bm{q}} h^{v{\rm (s)}}_{{\rm BD}\bm{q}} h^{xy{\rm (as)}}_{{\rm CD}\bm{q}}
+h^{\perp{\rm (s)}}_{{\rm BC}\bm{q}} h^{v{\rm (s)}}_{{\rm CD}\bm{q}} h^{xy{\rm (as)}}_{{\rm BD}\bm{q}}
+h^{\perp{\rm (s)}}_{{\rm BC}\bm{q}} h^{xy{\rm (s)}}_{{\rm AB}\bm{q}} h^{v{\rm (as)}}_{{\rm AC}\bm{q}}
+h^{\perp{\rm (s)}}_{{\rm BC}\bm{q}} h^{xy{\rm (s)}}_{{\rm AC}\bm{q}} h^{v{\rm (as)}}_{{\rm AB}\bm{q}} \nonumber \\
&-h^{\perp{\rm (s)}}_{{\rm BC}\bm{q}} h^{xy{\rm (s)}}_{{\rm BD}\bm{q}} h^{v{\rm (as)}}_{{\rm CD}\bm{q}}
-h^{\perp{\rm (s)}}_{{\rm BC}\bm{q}} h^{xy{\rm (s)}}_{{\rm CD}\bm{q}} h^{v{\rm (as)}}_{{\rm BD}\bm{q}}
-h^{\perp{\rm (s)}}_{{\rm BD}\bm{q}} h^{v{\rm (s)}}_{{\rm AB}\bm{q}} h^{xy{\rm (as)}}_{{\rm AD}\bm{q}}
-h^{\perp{\rm (s)}}_{{\rm BD}\bm{q}} h^{v{\rm (s)}}_{{\rm AD}\bm{q}} h^{xy{\rm (as)}}_{{\rm AB}\bm{q}}
-h^{\perp{\rm (s)}}_{{\rm BD}\bm{q}} h^{v{\rm (s)}}_{{\rm BC}\bm{q}} h^{xy{\rm (as)}}_{{\rm CD}\bm{q}} \nonumber \\
&+h^{\perp{\rm (s)}}_{{\rm BD}\bm{q}} h^{v{\rm (s)}}_{{\rm CD}\bm{q}} h^{xy{\rm (as)}}_{{\rm BC}\bm{q}}
+h^{\perp{\rm (s)}}_{{\rm BD}\bm{q}} h^{xy{\rm (s)}}_{{\rm AB}\bm{q}} h^{v{\rm (as)}}_{{\rm AD}\bm{q}}
+h^{\perp{\rm (s)}}_{{\rm BD}\bm{q}} h^{xy{\rm (s)}}_{{\rm AD}\bm{q}} h^{v{\rm (as)}}_{{\rm AB}\bm{q}}
+h^{\perp{\rm (s)}}_{{\rm BD}\bm{q}} h^{xy{\rm (s)}}_{{\rm BC}\bm{q}} h^{v{\rm (as)}}_{{\rm CD}\bm{q}}
-h^{\perp{\rm (s)}}_{{\rm BD}\bm{q}} h^{xy{\rm (s)}}_{{\rm CD}\bm{q}} h^{v{\rm (as)}}_{{\rm BC}\bm{q}} \nonumber \\
&-h^{\perp{\rm (s)}}_{{\rm CD}\bm{q}} h^{v{\rm (s)}}_{{\rm AC}\bm{q}} h^{xy{\rm (as)}}_{{\rm AD}\bm{q}}
-h^{\perp{\rm (s)}}_{{\rm CD}\bm{q}} h^{v{\rm (s)}}_{{\rm AD}\bm{q}} h^{xy{\rm (as)}}_{{\rm AC}\bm{q}}
-h^{\perp{\rm (s)}}_{{\rm CD}\bm{q}} h^{v{\rm (s)}}_{{\rm BC}\bm{q}} h^{xy{\rm (as)}}_{{\rm BD}\bm{q}}
-h^{\perp{\rm (s)}}_{{\rm CD}\bm{q}} h^{v{\rm (s)}}_{{\rm BD}\bm{q}} h^{xy{\rm (as)}}_{{\rm BC}\bm{q}}
+h^{\perp{\rm (s)}}_{{\rm CD}\bm{q}} h^{xy{\rm (s)}}_{{\rm AC}\bm{q}} h^{v{\rm (as)}}_{{\rm AD}\bm{q}} \nonumber \\
&+h^{\perp{\rm (s)}}_{{\rm CD}\bm{q}} h^{xy{\rm (s)}}_{{\rm AD}\bm{q}} h^{v{\rm (as)}}_{{\rm AC}\bm{q}}
+h^{\perp{\rm (s)}}_{{\rm CD}\bm{q}} h^{xy{\rm (s)}}_{{\rm BC}\bm{q}} h^{v{\rm (as)}}_{{\rm BD}\bm{q}}
+h^{\perp{\rm (s)}}_{{\rm CD}\bm{q}} h^{xy{\rm (s)}}_{{\rm BD}\bm{q}} h^{v{\rm (as)}}_{{\rm BC}\bm{q}}
+h^{v{\rm (s)}}_{{\rm AB}\bm{q}} h^{xy{\rm (s)}}_{{\rm AC}\bm{q}} h^{\perp{\rm (as)}}_{{\rm BC}\bm{q}}
+h^{v{\rm (s)}}_{{\rm AB}\bm{q}} h^{xy{\rm (s)}}_{{\rm AD}\bm{q}} h^{\perp{\rm (as)}}_{{\rm BD}\bm{q}} \nonumber \\
&+h^{v{\rm (s)}}_{{\rm AB}\bm{q}} h^{xy{\rm (s)}}_{{\rm BC}\bm{q}} h^{\perp{\rm (as)}}_{{\rm AC}\bm{q}}
+h^{v{\rm (s)}}_{{\rm AB}\bm{q}} h^{xy{\rm (s)}}_{{\rm BD}\bm{q}} h^{\perp{\rm (as)}}_{{\rm AD}\bm{q}}
-h^{v{\rm (s)}}_{{\rm AC}\bm{q}} h^{xy{\rm (s)}}_{{\rm AB}\bm{q}} h^{\perp{\rm (as)}}_{{\rm BC}\bm{q}}
+h^{v{\rm (s)}}_{{\rm AC}\bm{q}} h^{xy{\rm (s)}}_{{\rm AD}\bm{q}} h^{\perp{\rm (as)}}_{{\rm CD}\bm{q}}
+h^{v{\rm (s)}}_{{\rm AC}\bm{q}} h^{xy{\rm (s)}}_{{\rm BC}\bm{q}} h^{\perp{\rm (as)}}_{{\rm AB}\bm{q}} \nonumber \\
&+h^{v{\rm (s)}}_{{\rm AC}\bm{q}} h^{xy{\rm (s)}}_{{\rm CD}\bm{q}} h^{\perp{\rm (as)}}_{{\rm AD}\bm{q}}
-h^{v{\rm (s)}}_{{\rm AD}\bm{q}} h^{xy{\rm (s)}}_{{\rm AB}\bm{q}} h^{\perp{\rm (as)}}_{{\rm BD}\bm{q}}
-h^{v{\rm (s)}}_{{\rm AD}\bm{q}} h^{xy{\rm (s)}}_{{\rm AC}\bm{q}} h^{\perp{\rm (as)}}_{{\rm CD}\bm{q}}
+h^{v{\rm (s)}}_{{\rm AD}\bm{q}} h^{xy{\rm (s)}}_{{\rm BD}\bm{q}} h^{\perp{\rm (as)}}_{{\rm AB}\bm{q}}
+h^{v{\rm (s)}}_{{\rm AD}\bm{q}} h^{xy{\rm (s)}}_{{\rm CD}\bm{q}} h^{\perp{\rm (as)}}_{{\rm AC}\bm{q}} \nonumber \\
&-h^{v{\rm (s)}}_{{\rm BC}\bm{q}} h^{xy{\rm (s)}}_{{\rm AB}\bm{q}} h^{\perp{\rm (as)}}_{{\rm AC}\bm{q}}
-h^{v{\rm (s)}}_{{\rm BC}\bm{q}} h^{xy{\rm (s)}}_{{\rm AC}\bm{q}} h^{\perp{\rm (as)}}_{{\rm AB}\bm{q}}
+h^{v{\rm (s)}}_{{\rm BC}\bm{q}} h^{xy{\rm (s)}}_{{\rm BD}\bm{q}} h^{\perp{\rm (as)}}_{{\rm CD}\bm{q}}
+h^{v{\rm (s)}}_{{\rm BC}\bm{q}} h^{xy{\rm (s)}}_{{\rm CD}\bm{q}} h^{\perp{\rm (as)}}_{{\rm BD}\bm{q}}
-h^{v{\rm (s)}}_{{\rm BD}\bm{q}} h^{xy{\rm (s)}}_{{\rm AB}\bm{q}} h^{\perp{\rm (as)}}_{{\rm AD}\bm{q}} \nonumber \\
&-h^{v{\rm (s)}}_{{\rm BD}\bm{q}} h^{xy{\rm (s)}}_{{\rm AD}\bm{q}} h^{\perp{\rm (as)}}_{{\rm AB}\bm{q}}
-h^{v{\rm (s)}}_{{\rm BD}\bm{q}} h^{xy{\rm (s)}}_{{\rm BC}\bm{q}} h^{\perp{\rm (as)}}_{{\rm CD}\bm{q}}
+h^{v{\rm (s)}}_{{\rm BD}\bm{q}} h^{xy{\rm (s)}}_{{\rm CD}\bm{q}} h^{\perp{\rm (as)}}_{{\rm BC}\bm{q}}
-h^{v{\rm (s)}}_{{\rm CD}\bm{q}} h^{xy{\rm (s)}}_{{\rm AC}\bm{q}} h^{\perp{\rm (as)}}_{{\rm AD}\bm{q}}
-h^{v{\rm (s)}}_{{\rm CD}\bm{q}} h^{xy{\rm (s)}}_{{\rm AD}\bm{q}} h^{\perp{\rm (as)}}_{{\rm AC}\bm{q}} \nonumber \\
&-h^{v{\rm (s)}}_{{\rm CD}\bm{q}} h^{xy{\rm (s)}}_{{\rm BC}\bm{q}} h^{\perp{\rm (as)}}_{{\rm BD}\bm{q}}
-h^{v{\rm (s)}}_{{\rm CD}\bm{q}} h^{xy{\rm (s)}}_{{\rm BD}\bm{q}} h^{\perp{\rm (as)}}_{{\rm BC}\bm{q}}
+h^{\perp{\rm (as)}}_{{\rm AB}\bm{q}} h^{v{\rm (as)}}_{{\rm AC}\bm{q}} h^{xy{\rm (as)}}_{{\rm BC}\bm{q}}
+h^{\perp{\rm (as)}}_{{\rm AB}\bm{q}} h^{v{\rm (as)}}_{{\rm AD}\bm{q}} h^{xy{\rm (as)}}_{{\rm BD}\bm{q}}
-h^{\perp{\rm (as)}}_{{\rm AB}\bm{q}} h^{v{\rm (as)}}_{{\rm BC}\bm{q}} h^{xy{\rm (as)}}_{{\rm AC}\bm{q}} \nonumber \\
&-h^{\perp{\rm (as)}}_{{\rm AB}\bm{q}} h^{v{\rm (as)}}_{{\rm BD}\bm{q}} h^{xy{\rm (as)}}_{{\rm AD}\bm{q}}
-h^{\perp{\rm (as)}}_{{\rm AC}\bm{q}} h^{v{\rm (as)}}_{{\rm AB}\bm{q}} h^{xy{\rm (as)}}_{{\rm BC}\bm{q}}
+h^{\perp{\rm (as)}}_{{\rm AC}\bm{q}} h^{v{\rm (as)}}_{{\rm AD}\bm{q}} h^{xy{\rm (as)}}_{{\rm CD}\bm{q}}
+h^{\perp{\rm (as)}}_{{\rm AC}\bm{q}} h^{v{\rm (as)}}_{{\rm BC}\bm{q}} h^{xy{\rm (as)}}_{{\rm AB}\bm{q}}
-h^{\perp{\rm (as)}}_{{\rm AC}\bm{q}} h^{v{\rm (as)}}_{{\rm CD}\bm{q}} h^{xy{\rm (as)}}_{{\rm AD}\bm{q}} \nonumber \\
&-h^{\perp{\rm (as)}}_{{\rm AD}\bm{q}} h^{v{\rm (as)}}_{{\rm AB}\bm{q}} h^{xy{\rm (as)}}_{{\rm BD}\bm{q}}
-h^{\perp{\rm (as)}}_{{\rm AD}\bm{q}} h^{v{\rm (as)}}_{{\rm AC}\bm{q}} h^{xy{\rm (as)}}_{{\rm CD}\bm{q}}
+h^{\perp{\rm (as)}}_{{\rm AD}\bm{q}} h^{v{\rm (as)}}_{{\rm BD}\bm{q}} h^{xy{\rm (as)}}_{{\rm AB}\bm{q}}
+h^{\perp{\rm (as)}}_{{\rm AD}\bm{q}} h^{v{\rm (as)}}_{{\rm CD}\bm{q}} h^{xy{\rm (as)}}_{{\rm AC}\bm{q}}
+h^{\perp{\rm (as)}}_{{\rm BC}\bm{q}} h^{v{\rm (as)}}_{{\rm AB}\bm{q}} h^{xy{\rm (as)}}_{{\rm AC}\bm{q}} \nonumber \\
&-h^{\perp{\rm (as)}}_{{\rm BC}\bm{q}} h^{v{\rm (as)}}_{{\rm AC}\bm{q}} h^{xy{\rm (as)}}_{{\rm AB}\bm{q}}
+h^{\perp{\rm (as)}}_{{\rm BC}\bm{q}} h^{v{\rm (as)}}_{{\rm BD}\bm{q}} h^{xy{\rm (as)}}_{{\rm CD}\bm{q}}
-h^{\perp{\rm (as)}}_{{\rm BC}\bm{q}} h^{v{\rm (as)}}_{{\rm CD}\bm{q}} h^{xy{\rm (as)}}_{{\rm BD}\bm{q}}
+h^{\perp{\rm (as)}}_{{\rm BD}\bm{q}} h^{v{\rm (as)}}_{{\rm AB}\bm{q}} h^{xy{\rm (as)}}_{{\rm AD}\bm{q}}
-h^{\perp{\rm (as)}}_{{\rm BD}\bm{q}} h^{v{\rm (as)}}_{{\rm AD}\bm{q}} h^{xy{\rm (as)}}_{{\rm AB}\bm{q}} \nonumber \\
&-h^{\perp{\rm (as)}}_{{\rm BD}\bm{q}} h^{v{\rm (as)}}_{{\rm BC}\bm{q}} h^{xy{\rm (as)}}_{{\rm CD}\bm{q}}
+h^{\perp{\rm (as)}}_{{\rm BD}\bm{q}} h^{v{\rm (as)}}_{{\rm CD}\bm{q}} h^{xy{\rm (as)}}_{{\rm BC}\bm{q}}
+h^{\perp{\rm (as)}}_{{\rm CD}\bm{q}} h^{v{\rm (as)}}_{{\rm AC}\bm{q}} h^{xy{\rm (as)}}_{{\rm AD}\bm{q}}
-h^{\perp{\rm (as)}}_{{\rm CD}\bm{q}} h^{v{\rm (as)}}_{{\rm AD}\bm{q}} h^{xy{\rm (as)}}_{{\rm AC}\bm{q}}
+h^{\perp{\rm (as)}}_{{\rm CD}\bm{q}} h^{v{\rm (as)}}_{{\rm BC}\bm{q}} h^{xy{\rm (as)}}_{{\rm BD}\bm{q}} \nonumber \\
&-h^{\perp{\rm (as)}}_{{\rm CD}\bm{q}} h^{v{\rm (as)}}_{{\rm BD}\bm{q}} h^{xy{\rm (as)}}_{{\rm BC}\bm{q}})
\end{align}
where $H'_{3\bm{q}}$ and $H'_{4\bm{q}}$ are obtained by replacing the superscript $\perp$ in $H'_{2\bm{q}}$ with $v$ and $xy$, respectively, and multiplying $-1$.

\end{widetext}

\begin{acknowledgments}
This research was supported by JSPS KAKENHI Grants Numbers JP19K03752, JP19H01834, JP21H01037, and by JST PRESTO (JPMJPR20L8). 
Parts of the numerical calculations were performed in the supercomputing systems in ISSP, the University of Tokyo.
\end{acknowledgments}

\bibliographystyle{apsrev}
\bibliography{ref}

\end{document}